\documentclass[twocolumn,trackchanges]{aastex61}
\hypersetup{bookmarksnumbered=true}

\usepackage{amsmath}
\usepackage{amsfonts}
\usepackage{amssymb}
\usepackage{apjfonts}
\allowdisplaybreaks[1]

\renewcommand{\vr}{v_{\mathrm{r}}}
\newcommand{\vt}{v_{\mathrm{t}}}
\newcommand{\vrt}{v_{\mathrm{r(t)}}}
\newcommand{\vlos}{v_{\mathrm{los}}}

\newcommand{\mh}{M_{\mathrm{h}}}
\newcommand{\rh}{R_{\mathrm{h}}}

\newcommand{\msun}{M_{\odot}}
\newcommand{\lsun}{L_{\odot}}
\newcommand{\gyr}{\mathrm{Gyr}}
\newcommand{\mpc}{\mathrm{Mpc}}
\newcommand{\kpc}{\mathrm{kpc}}
\newcommand{\kms}{\mathrm{km \, s}^{-1}}
\newcommand{\msunh}{h^{-1} \, M_{\odot}}
\newcommand{\mpch}{h^{-1} \, \mathrm{Mpc}}
\newcommand{\kpch}{h^{-1} \, \mathrm{kpc}}
\newcommand{\masyr}{\mathrm{mas \, yr}^{-1}}
\newcommand{\mathd}{\mathrm{d}}

\newcommand{\avg}[1]{\left\langle #1 \right\rangle}

\newcommand{\LCDM}{\Lambda\mathrm{CDM}}
\newcommand{\pmra}{\mu_\alpha}
\newcommand{\pmdec}{\mu_\delta}

\newcommand{\mesti}{M_{\mathrm{esti}}}
\newcommand{\mcorr}{\hat{M}_{\mathrm{esti}}}
\newcommand{\mtrue}{M_{\mathrm{true}}}
\newcommand{\avgmesti}{\overline{M}_{\mathrm{esti}}}
\newcommand{\avgmcorr}{\hat{\overline{M}}_{\mathrm{esti}}}
\newcommand{\bias}{\eta}

\newcommand{\refsec}[1]{Section~\ref{sec:#1}}
\newcommand{\reffig}[1]{Figure~\ref{fig:#1}}

\newcommand{\refeq}[1]{Equation~(\ref{eq:#1})}

\newcommand{\figdir}{.}

\received{xxxx x, 2017}
\revised{xxxx x, 2017}
\accepted{xxxx x, 2017}
\submitjournal{ApJ}

\shorttitle{Determination of Dark Matter Halo Mass}
\shortauthors{Li et al.}

\begin{document}

\title{Determination of Dark Matter Halo Mass from Dynamics of Satellite Galaxies}

\correspondingauthor{Y.P. Jing}
\email{lizz@shao.ac.cn, ypjing@sjtu.edu.cn}

\author{Zhao-Zhou Li}
\affil{Key Laboratory for Research in Galaxies and Cosmology, Shanghai Astronomical Observatory, 80 Nandan Road, Shanghai 200030, China}
\affil{University of the Chinese Academy of Sciences, No.19A Yuquan Road, Beijing 100049, China}

\author{Y.P. Jing}
\affil{Center for Astronomy and Astrophysics, School of Physics and
  Astronomy, Shanghai Jiao Tong University, 955 Jianchuan Road,
  Shanghai 200240, China}
\affil{IFSA Collaborative Innovation Center, Shanghai Jiao
  Tong University, Shanghai 200240, China}

\author{Yong-Zhong Qian}
\affil{School of Physics and Astronomy, University of Minnesota, Minneapolis, MN 55455, USA}
\affil{Tsung-Dao Lee Institute, Shanghai 200240, China}

\author{Zhen Yuan}
\affil{Center for Astronomy and Astrophysics, School of Physics and Astronomy, Shanghai Jiao Tong University, 955 Jianchuan Road, Shanghai 200240, China}

\author{Dong-Hai Zhao}
\affil{Key Laboratory for Research in Galaxies and Cosmology, Shanghai Astronomical Observatory, 80 Nandan Road, Shanghai 200030, China}


\begin{abstract}

We show that the mass of a dark matter halo can be inferred from the dynamical status
of its satellite galaxies. Using 9 dark-matter simulations of halos like the Milky Way (MW), 
we find that the present-day substructures in each halo follow a characteristic distribution 
in the phase space of orbital binding energy and angular momentum, and that this distribution 
is similar from halo to halo but has an intrinsic dependence on the halo formation history.
We construct this distribution directly from the simulations for a specific halo and extend the 
result to halos of similar formation history but different masses by scaling. The mass of an 
observed halo can then be estimated by maximizing the likelihood in comparing the measured 
kinematic parameters of its satellite galaxies with these distributions. We test the validity and 
accuracy of this method with mock samples taken from the simulations. Using the positions,
radial velocities, and proper motions of 9 tracers and assuming observational uncertainties
comparable to those of MW satellite galaxies, we find that the halo mass can be recovered 
to within $\sim 40$\%. The accuracy can be improved to within $\sim$25\% if 30 tracers are
used. However, the dependence of the phase-space distribution on the halo formation history
sets a minimum uncertainty of $\sim 20$\% that cannot be reduced by
using more tracers. 
We believe that this minimum uncertainty also applies to any mass 
determination for a halo when the phase space information of other kinematic tracers is used.

\end{abstract}



\section{Introduction} \label{sec:intro}
We present a method to estimate the mass of a dark matter halo from the
dynamical status of its satellite galaxies.
In the framework of hierarchical structure formation based on the concordance 
cold dark matter ($\LCDM$) cosmology, the mass of a dark matter halo is closely
related to its many other properties such as structure, dynamics, and formation history. 
In the case of the Milky Way (MW), a number of theoretical predictions or interpretations 
of observations, for example, the baryon fraction \citep[e.g.][]{Zaritsky2017} and
the problem of missing massive satellites 
\citep[e.g.][]{Boylan-Kolchin2011, Wang2012b, Cautun2014}, 
depend on the MW halo mass. Various methods have been proposed to measure 
this important quantity (see \citealt{Courteau2014,Bland-Hawthorn2016a} 
for reviews and \citealt{Wang2015b} for a comparison of recent measurements). 
Though these measurements are roughly consistent, they result in a factor of
$\sim 3$ difference in the estimated MW halo mass. The scatter might be even larger 
if systematic uncertainties are included \citep{Han2016a, Wang2016b}. 
Clearly, there is a need for more accurate methods to determine the MW halo mass.

The MW halo mass can be constrained by the abundances of certain constituents, 
such as the baryon fraction \citep{Zaritsky2017}, the total stellar mass \citep{Guo2011}, 
and the number of satellite galaxies above a specific threshold 
\citep[e.g.][]{Starkenburg2013, Rodriguez-Puebla2013, Cautun2014}.
Timing argument is widely used to give another mass estimator by modeling the 
expansion of the Local Group galaxies \citep{Kahn1959,Li2008,Penarrubia2016,Banik2016}. 
Perhaps the most powerful and direct method to estimate the MW halo mass is to
use dynamical tracers. In this regard, the mass distribution within $\sim 100$ kpc is 
reasonably well constrained by the kinematics of stars \citep[e.g.][]{Xue2008, Huang2016} 
or a stellar stream \citep[e.g.][]{Gibbons2014}. However, due to the limited spatial 
distribution of these tracers, extrapolation is needed to obtain the total halo mass, which
often depends on the assumed parametric form for the overall density profile.

The outer region of the MW can be investigated more directly by using its
satellite galaxies, which lie far beyond the other tracers. However, this approach 
was limited for a long time by the small sample size, large uncertainties
in distance estimates, and lack of proper-motion measurement. Fortunately,
both the sample size and precision of distance measurement have increased
greatly over the past decade (see \citealt{McConnachie2012} 
for a recent compilation of observations\footnote{An updated compilation can be 
downloaded from \url{http://www.astro.uvic.ca/\~alan/Nearby_Dwarf_Database.html}}).
In addition, with the unprecedented precision of the HST and the new generation of 
ground-based telescopes, proper motions of bright satellites have been measured
(e.g. \citealt{Piatek2002, Piatek2003, Piatek2005}, also see Table 1 of 
\citealt{Pawlowski2013a} for a summary of currently available measurements).
Consequently, such satellites are fully characterized in the 6D phase space of
position and velocity and their orbits can be computed assuming a potential. 
Currently, proper motions are available for 12 of the 13 satellite galaxies (the 
exception being Canes Venatici I) that are more luminous than $10^5\lsun$ 
and within 300 kpc from the MW center.

However, unlike stellar tracers, the limited number of satellite galaxies does not 
allow direct calculation of the velocity dispersion profile or rotation curve. Instead, 
analytical models of the dynamical status or comparisons with numerical simulations
are required. Previous studies using satellite galaxies considered the orbital energy 
of Leo I \citep{Boylan-Kolchin2013}, velocity moments \citep{Watkins2010a}, 
orbital ellipticity distribution \citep{Barber2014}, and probability distribution of 
orbital parameters \citep{Eadie2015,Eadie2016}. These approaches encounter
several difficulties. When the density and velocity anisotropy profiles of the tracer 
population are assumed, the inferred mass distribution depends sensitively on 
the assumptions \citep[e.g.][]{Watkins2010a,Eadie2016}, which requires further
systematic study. In addition, analytical methods assume that all satellites are 
bound in a steady state with random orbital phases, which may not hold for
all halos. The influence of deviations from a steady state and halo-to-halo scatter 
has yet to be taken into full consideration. Another difficulty is how to treat 
observational errors properly, as the measurement uncertainty differs substantially 
from satellite to satellite. In this paper we develop a new method that either avoids
or addresses the above issues in using satellite galaxies to estimate the MW halo 
mass.

We base our method on dark-matter simulations of MW-like halos and
associate satellite galaxies with subhalos of a simulated halo.
We construct the distribution of subhalos in the phase space of orbital binding 
energy and angular momentum directly from the simulations without assuming 
a steady state or any particular form of velocity anisotropy. We also take 
into account observational uncertainties of satellite galaxies.
We estimate the halo mass by maximizing the likelihood in comparing the 
observed orbital parameters of satellite galaxies with the phase-space 
distribution derived from simulations. We test the validity of this method and 
investigate its systematics using mock samples taken from simulations.
We also study the dependence of this halo mass estimator on observational 
uncertainties, the number of satellites used, and halo-to-halo scatter.
While our method is motivated by improving the estimate of the MW halo
mass, it can be extended to other MW-like halos as well.

The plan of this paper is as follows. 
We outline our method in \refsec{method} and show how to construct the 
phase-space distribution of subhalos from simulations in \refsec{template}.
We discuss systematic tests by mock samples in \refsec{test} and give
conclusions in \refsec{conclusion}.

\section{Method} \label{sec:method}

Our basic assumption is that for a present-day halo of mass $\mh$, 
its substructures have a characteristic distribution $p(E,L|\mh)$ in the phase space
of orbital binding energy $E$ and angular momentum $L$ (see below for definition).
Then the unknown mass of a halo can be inferred by comparing the observed orbital 
parameters of its substructure tracers with the phase-space distributions derived 
from simulations for different $\mh$. In practice, dwarf satellite galaxies are the 
outmost tracers for the MW. To develop the halo mass estimator, we consider
such satellites as a subset of the surviving subhalos for a halo 
in terms of kinematics\footnote{
    We assume that the orbits of satellites are not subject to significant selection effects.
    Satellite samples are usually selected by some luminosity threshold. Using the data
    from \citet{McConnachie2012}, we have checked that both the space and radial 
    velocity distributions of the most luminous ($>10^5 \lsun$) 13 MW satellites
    agree with those of the nearly complete sample of $\sim 25$ fainter ($>10^4 \lsun$)
    satellites. This result is consistent with our assumption.
}.
Hereafter, the simulated halo 
that provides the calculated phase-space distribution is referred to as the 
\textit{template halo}. The halo whose mass is to be determined is referred to as the 
\textit{test halo}.

We characterize the orbit in the potential of a halo by the corresponding binding 
energy $E$ and angular momentum $L$ per unit mass. Specifically,
\begin{equation}
\begin{split}
&E = - \Phi (r) - \frac{1}{2} (\vr^2 + \vt^2), \\
&L = r\vt,
\end{split}
\end{equation}
where $r$, $\vr$, and $\vt$ are the distance, the radial and tangential velocity 
relative to the center of the host halo, respectively, and 
\begin{equation}
\Phi (r) = - \int_r^{r_0} \frac{G M_{\Delta} (r')}{r'^2} \mathd r'
\end{equation}
is the gravitational potential. In the above equation, $r_0$ corresponds to
the zero potential point, $G$ is the gravitational constant, and
\begin{equation}
M_{\Delta} (r) = \int_0^r 4 \pi [\rho (r') - \bar{\rho}] r'^2 \mathd r'
\end{equation}
represents the mass exceeding the mean cosmic background,
where $\rho(r)$ is the dark matter density profile of the host halo and
$\bar{\rho}$ is the mean cosmic density. We adopt $r_0 = 1 \mpch$
and have checked that using $r_0=3 \mpch$ instead makes little difference 
in the results. 

There are two reasons why we do not use the observable parameters $r$, $\vr$, 
and $\vt$ directly although this alternative seems to provide more information.
First, the $E$ and $L$ of a subhalo are approximately conserved after its
infall into the host halo. They are less mixed in phase space over time 
and also less sensitive to individual merger events that produce 
halo-to-halo scatter. Second, due to the finite number of subhalos in the 
simulations, the constructed phase space of 
$r$, $\vr$, and $\vt$ is more sparse, and therefore, more discontinuous.
This problem is mitigated by using the phase space of the corresponding
$E$ and $L$ instead. Hereafter, ``phase space'' means $E$-$L$ space.

For a template halo of mass $\mh$, we construct the phase-space distribution
$p(E,L|\mh)$ directly from the simulations. Specifically,
\begin{equation}
p(E, L|\mh) = \frac{1}{n_{\mathrm{sub}}} \sum_{i = 1}^{n_{\mathrm{sub}}} 
\tilde p (E, L| \mathrm{sub}_i),
\label{eq:p(E,L)}
\end{equation}
where $n_{\mathrm{sub}}$ is the number of selected subhalos,
and $\tilde p (E, L| \mathrm{sub}_i)$ represents the probability density for the $i$-th 
subhalo to be ``observed'' at $(E,L)$. As described in detail in \refsec{template},
$\tilde p (E, L| \mathrm{sub}_i)$ serves as the kernel function in the kernel density 
estimation to transform the discrete distribution of subhalos 
in phase space into a continuous one.
																				
The utility of template halos is greatly extended by the scaling
technique. For the mass range of our interest for the Milky
  Way halo,  dark-matter halos are built up approximately in a self-similar manner, 
thus we can scale a halo to a different mass 
while keeping the formation history and relaxation status unchanged.
Specifically, the distribution $p(E,L|\mh')$ for a halo of mass $\mh'$ 
can be obtained from $p(E,L|\mh)$ by using 
\begin{equation}
\label{eq:scaling}
\begin{split}
r' = (\mh' / \mh)^{1 / 3} r ,\\
\vrt' = (\mh' / \mh)^{1 / 3} \vrt, \\
\Phi' = (\mh' / \mh)^{2 / 3} \Phi,
\end{split}
\end{equation}
for each of the subhalos in the halo of mass $\mh$.
The subhalo mass is scaled as $m'=(\mh' / \mh) m$.
In this way, we can construct a family of distributions $p(E,L|\mh')$
for a range of halo mass $\mh'$ from a single template halo.

To infer the unknown mass of a test halo hosting a set of satellites with
observed $(r,v_r,v_t)$, we calculate $(E,L)$ for each satellite using the
potential of a scaled template halo of mass $\mh'$ and further compute 
the likelihood
\begin{equation}
\mathcal{L} (\mathrm{obs} |M_h') = \prod^{N_{\mathrm{sat}}}_{k = 1} p(E_k, L_k |M_h'),
\label{eq:likelihood}
\end{equation}
where $N_{\mathrm{sat}}$ is the number of observed satellites, and
$(E_k, L_k)$ correspond to the $k$-th satellite. The likelihood
$\mathcal{L} (\mathrm{obs} |\mh')$ can be calculated for a range of 
template halos scaled to different $\mh'$.
Assuming that the test halo has the same formation history and relaxation 
status as the template halos, we can infer the unknown mass of the test halo
by maximizing $\mathcal{L} (\mathrm{obs} |\mh')$, which gives
the Maximum Likelihood Estimator (MLE) for the mass
\begin{equation}
M_{\mathrm{esti}} = \arg \max_{\mh'} \mathcal{L} (\mathrm{obs} |\mh').
\end{equation}

\begin{figure*}[ht!]
\epsscale{1.2}
\plotone{\figdir/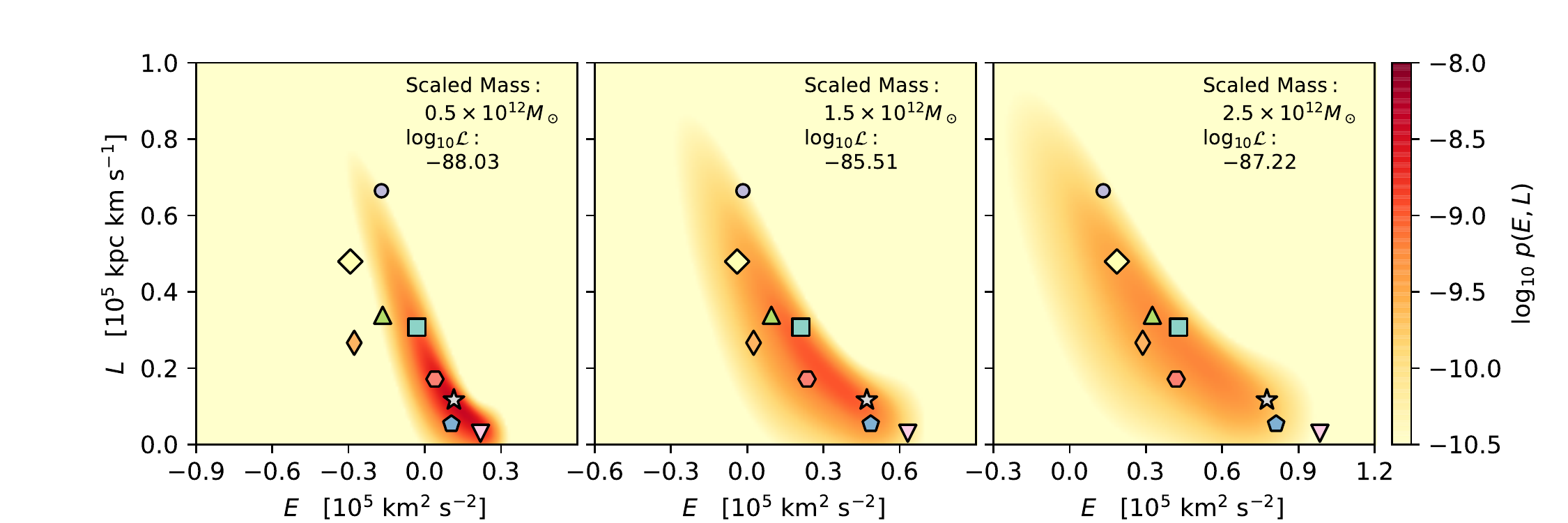}
\caption{Comparison of mock observations with the phase-space 
distributions constructed from simulations of template halo A1 
scaled to different halo masses. Template halo A1 is also used as
a test halo and its most massive 9 subhalos are chosen as
satellites for mock observations. Symbols with black border and
colored contours represent mock data and constructed phase-space 
distributions, respectively. The scaled halo mass and the corresponding
likelihood of mock observations are shown in each panel. It can be seen 
that among the three cases, the likelihood is the largest for the middle 
one, whose halo mass is also closest to the true value of
$\approx 1.6\times 10^{12}\msun$.}
\label{fig:scaling}
\end{figure*}

The above method is illustrated by \reffig{scaling}, which shows
how $\mathcal{L} (\mathrm{obs} |\mh')$ changes with $\mh'$.
Template halo A1 in our simulations, which has 
a true mass of $M_h\approx 1.6\times 10^{12}\msun$, is
also used as a test halo and its most massive 9
subhalos are chosen as satellites, for which mock observations of
$(r,v_r,v_t)$ are made with the fiducial measurement precision
(see \refsec{observation} and \refsec{mock}).
The colored contours in \reffig{scaling} represent the phase-space 
distribution constructed from template halo A1 by scaling it to
$\mh'=0.5\times 10^{12}$, $1.5\times 10^{12}$, and
$2.5\times 10^{12}\msun$, respectively.
The symbols stand for the mock data on $(E,L)$ for the satellites.
Note that the ``observed'' $(r,v_r,v_t)$ for each satellite do not change
during scaling. Therefore, the mock data on $L$ remain the same but 
those on $E$ change with the potential of the scaled template halo. 
The observation points become more bound as $\mh'$ increases.
It can be seen from \reffig{scaling} that among the three $\mh'$ 
values, the likelihood of the observations is the largest for the middle 
one, which is also closest to the true value $\mh$.

Because the binding energy $E$ of a satellite depends on the template
halo mass $\mh'$, the likelihood $\mathcal{L} (\mathrm{obs} |\mh')$
cannot be converted in a straightforward manner into the probability 
distribution of the true halo mass even when the prior distribution of 
$\mh'$ is known. Nevertheless, we will show that the MLE $\mesti$ 
is indeed a good, though biased, indicator for the true halo mass.
Using Monte Carlo realization of mock samples, we find that the bias 
is approximately constant and define an average bias 
$\bias = \avg{\mesti/\mtrue}$ over the mock samples. Consequently, 
we obtain the bias-corrected estimator for the halo mass
\begin{equation}
\mcorr = \mesti / \bias.
\label{eq:correction}
\end{equation}

The above discussion assumes that the test and template halos have 
the same formation history and relaxation status. However, such information 
about the test halo is not readily available in practice. The lack of such
information then introduces an intrinsic uncertainty into our method.
We assess this uncertainty using 9 simulated halos with a wide range of 
formation history in \refsec{test}.
																				

\section{Construction of Subhalo Phase-Space Distribution} \label{sec:template}


Central to our method is the phase-space distribution $p(E,L|\mh)$
of subhalos for a template halo of mass $\mh$. This distribution is
constructed directly from our simulations taking into account
realistic observational uncertainties. The detailed procedure is 
described in this section.

\subsection{Simulations} \label{sec:simu}
In order to have enough substructures within a halo and resolve them 
with reasonable details,
we use the cosmological N-body simulation of \cite{Jing2002}
to select nine template halos for high-resolution resimulations. 
Each of these halos is required to be relatively isolated 
at redshift $z=0$\footnote{
The selection of relatively isolated halos is required by the zoom-in technique,  
because a close neighbor of low resolution may bring unpredictable 
numerical effects in re-simulations, while one of high resolution will 
consume too much computational time. We confirm in Appendix that 
the selection of relatively isolated halos does not affect our method.}
so that its distance to any more massive halo must exceed three times 
the sum of the virial radii of both halos. 
In addition, each template halo is required to have a mass of 
approximately $1.5\times 10^{12}M_{\odot}$ similar to that of the MW. 
The simulation was performed in a box of $100 \mpch$ on each side with a parallel
particle-particle-particle-mesh $\rm{P^3M}$ code using $512^3$ particles.
A $\LCDM$ cosmology was adopted with the density parameter $\Omega_m =
0.3$, the cosmological constant $\Omega_{\Lambda} = 0.7$, the Hubble
constant $h = 0.67$ in units of $100\ \kms \mpc^{-1}$, and the
slope $n_s=1$ and amplitude $\sigma_8 = 0.9$ of the primordial power
spectrum. While these parameters are not up to date, they are close
to the most recent results from the Planck mission. The differences in
the cosmological parameters have little effect on the conclusions of this study 
because our main concern is to develop a method of estimating halo masses
and test its validity.
For each template halo, we use the 
multiple-mass method to generate the initial conditions for 
zooming \citep{Jing2000} and carry out zoom-in resimulations using 
the public code Gadget2 \citep{Springel2005}. 
In the high-resolution region enclosing a template halo, these simulations have
a particle mass of $\sim 10^5 M_{\odot}$ (Table \ref{tab:simu}) and a softening length
of 0.15 $\kpch$. 

\begin{figure}[htb!]
\epsscale{1.2}
\plotone{\figdir/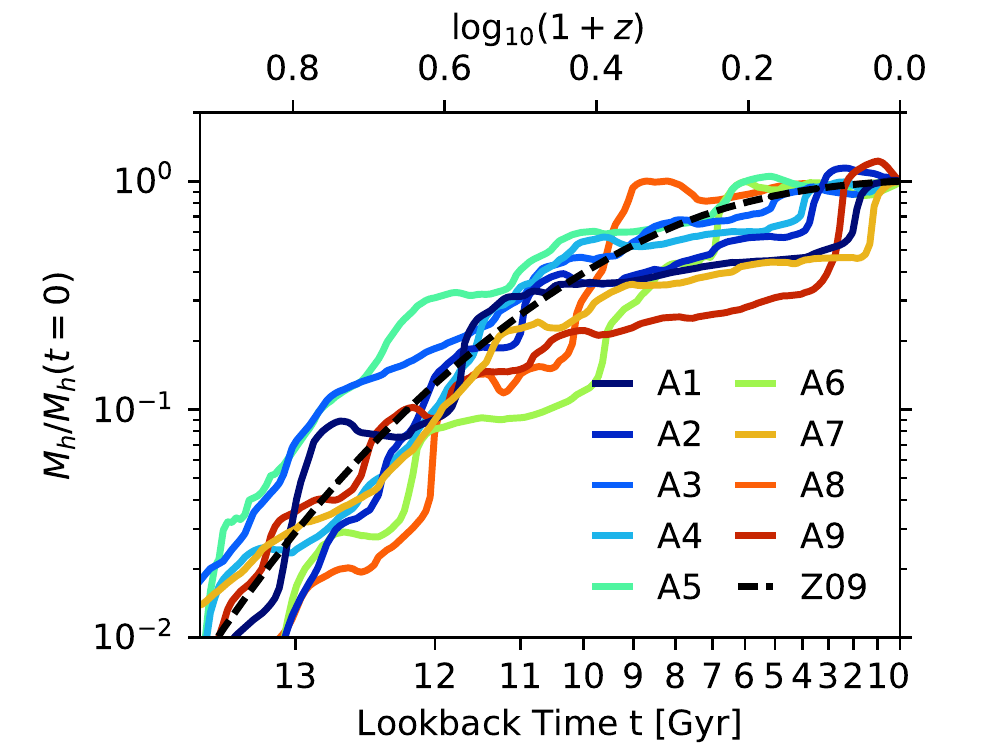}
\caption{
Growth history of template halos. The solid curves color-coded A1--A9
show the fraction of the present halo mass as a function of the lookback 
time $t$ for the corresponding halos. For reference, the dashed curve 
shows the median growth history for halos of $1.5 \times 10^{12}\msun$ in the 
model of \cite{Zhao2009}.
}
\label{fig:growth}
\end{figure}

We find halos using the standard Friends-of-Friends (FoF) algorithm with 
a linking length $b$ equal to 0.2 times the mean separation of high-resolution particles.
For ease of comparison with results in the literature, we define 
$\mh$ and $\rh$ as the mass and radius, respectively, of a spherical region with 
a mean density equal to 200 times the critical density of the universe.
The 9 template halos have $\mh \sim (1.3$--$1.6) \times 10^{12} M_{\odot}$ 
(Table \ref{tab:simu}) within $\rh \sim 230$~kpc at $z = 0$. 
As shown in \reffig{growth} and Table \ref{tab:simu}, these halos have
very different histories of mass growth, and therefore, cover a wide range of 
possible assembly history for an MW-like halo. Lacking the formation history 
of a test halo, we must resort to exploring a wide range of template halos to 
investigate the uncertainty from halo-to-halo scatter in our method of halo mass 
determination.

We use the Hierarchical Bound-Tracing (HBT) algorithm of \citet{Han2012} to identify
subhalos and build merger trees through time in the simulations. HBT
traces merger hierarchy of halos and subhalos with a physically-motivated 
unbinding algorithm, and thus has robust performance even in the dense inner 
region of a host halo, which fits our needs very well. 
The mass $m$ of a subhalo is defined to be its self-bound mass.
A subhalo can be identified if it contains at least 10 bound particles 
($\sim 10^6 M_{\odot}$). The positions and velocities of subhalos are
essential input to construction of their phase-space distribution. These
quantities are defined in HBT by the center of mass and bulk velocity of the
most bound 25\% of the particles in each subhalo (see \citealt{Han2012} for details).
The center of the largest subhalo is taken as the center of its host halo.

We have checked the completeness of subhalo samples in the high-resolution region of the zoom-in 
simulations. As expected \citep{Han2016}, within $2 R_h$, the number density 
profile of subhalos (including disrupted ones) in any given infall-mass bin coincides very well 
with the dark-matter density profile of the host halo. Therefore, the subhalo 
sample within $2 R_h$ is complete and not affected by the low-resolution particles.

\begin{deluxetable}{ccccc}
\tablecaption{Properties of Template Halos\label{tab:simu}}
\tablecolumns{5}
\tablewidth{0pt}
	\tablehead{
		\colhead{Halo} &
		\colhead{$m_p$} &
		\colhead{$M_h$} &
		\colhead{$t_{0.5}$} &
		\colhead{$t_{0.8}$}\\
		\colhead{} &
		\colhead{$(10^5 \msun)$} &
		\colhead{$(10^{12} \msun)$} &
		\colhead{$(\gyr)$} &
		\colhead{$(\gyr)$}
	}
	\startdata
	A1 & 0.99 & 1.58 &  3.14 & 1.79 \\
	A2 & 1.11 & 1.46 &  6.95 & 3.58 \\
	A3 & 0.96 & 1.61 &  9.21 & 5.06 \\
	A4 & 0.93 & 1.60 & 10.20 & 3.96 \\
	A5 & 0.96 & 1.60 & 10.55 & 6.65 \\
	A6 & 1.37 & 1.54 &  6.93 & 6.68 \\
	A7 & 1.02 & 1.55 &  1.42 & 1.09 \\
	A8 & 1.05 & 1.64 &  9.54 & 9.13 \\
	A9 & 0.92 & 1.38 &  2.71 & 2.46 \\
	\enddata
	\tablecomments{The columns are the particle mass $m_p$ 
	in the high-resolution region, the present ($z = 0$) mass $M_h$
	of the template halo, the lookback times $t_{0.5}$ and $t_{0.8}$ 
	when the halo first reached 50\% and 80\% of its present mass, 
	respectively.
	}
\end{deluxetable}

\subsection{Subhalo sample selection} \label{sec:subhalo_sample}

As satellite galaxies are intended as the subhalo tracers, we adopt the following
criteria to mimic these tracers in selecting the subhalos to construct 
the phase-space distribution for a template halo.
\begin{itemize}
	\item Maximum binding mass in history: $m_{\max} > 2 \times 10^{- 5} M_h$
	      ($\gtrsim 300$ particles)
	      	      	      	      	      	        
	      Subhalos containing only $\sim 10$ particles are vulnerable to numerical
	      instability.  As shown by \citet{Han2016}, at infall a subhalo 
          should be at least $\sim 30$ times more massive than the smallest resolved 
          subhalo to alleviate artificial disruptions.
          On the other hand, we would like to keep enough subhalos to
	      have good statistics. The above relatively low mass threshold is adopted as
	      a reasonable compromise. 
          We note that subhalos hosting the bright MW
	      dwarf galaxies were probably $\sim 100$ times larger than this limit at their 
	      infall. We will show that our method is not very sensitive to this mass selection.
	      	      	      	      	      	        
	\item Mass at $z = 0$: $m_0 \gtrsim 10^6 M_{\odot}$ ($> 10$ particles)
	      	      	      	      	      	        
	      This is a safe lower bound, as MW dwarf galaxies have high mass-to-light ratios 
	      \citep[e.g.][]{Wolf2010}, with the mass enclosed within 300 pc being
	      $\sim 10^7M_{\odot}$ for most of these satellites \citep{Strigari2008} and
	      that within the half-light radius being $\gtrsim 5\times 10^6 M_{\odot}$ for 
	      the classical dwarf galaxies \citep{McConnachie2012}.
          We have checked that doubling the lower bounds on $m_{\max}$ and $m_0$
          changes the results by only $\lesssim 5\%$.

	\item Distance to host halo center: $40\ \mathrm{kpc} < r < 300\ \mathrm{kpc}$
	      	      	      	      	      	        
	      This range covers the 9 MW satellites with adequate kinematic data and
	      excludes satellites experiencing strong tidal disruption due to extreme
	      proximity to the Galactic Center (GC) (see \refsec{observation}). 
	      As the absolute position of a subhalo changes with the scaled halo mass, 
	      this criterion makes the selected subhalo sample dependent on the scaling 
	      of a template halo. Consequently, to ensure completeness of the sample, 
	      a scaled template halo must satisfy $2 R_h>300\ \mathrm{kpc}$,
	      which limits the scaled halo mass to $M_h > 0.35\times 10^{12} M_{\odot}$.
	      This lower bound is below the expected MW halo mass and does not 
	      pose any limitation in practice.
\end{itemize}

\begin{figure*}[hbt!]
\epsscale{1.2}
\plotone{\figdir/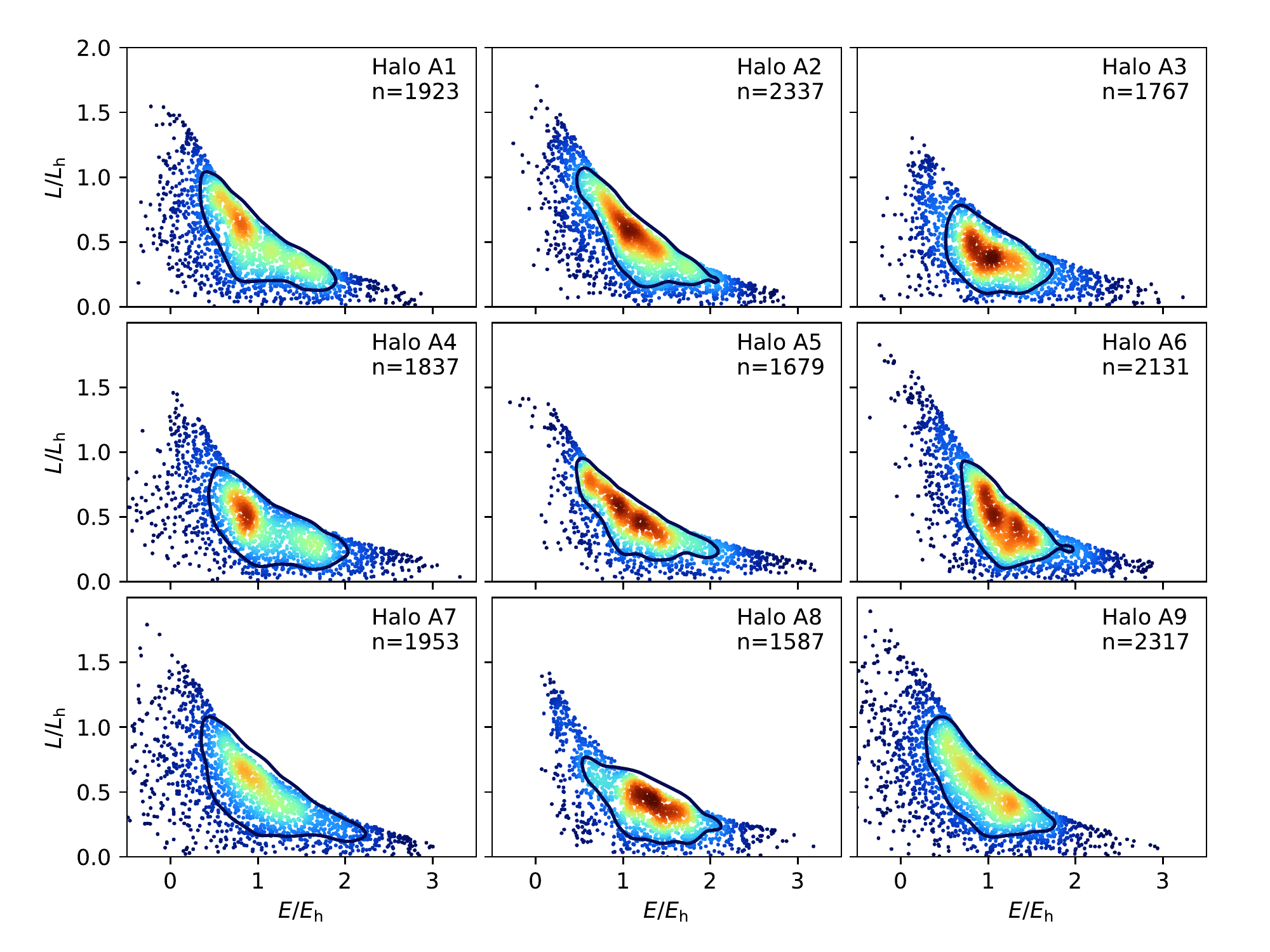}
\caption{Discrete phase-space distribution of subhalos for each of the template 
halos A1--A9. The units of $E$ and $L$ are $E_h=G M_h/R_h$ and
$L_h = \sqrt{G M_h R_h}$\,, respectively.
The number $n$ of subhalos in the selected sample is indicated 
for each halo. Each dot represents a subhalo and is colored according to the local 
number density in the phase space with red indicating higher density and the same 
color normalization for all halos. The curve is the equidensity contour enclosing 
half of the subhalos. The distributions show broad similarity but also clear 
differences.}
\label{fig:phasespace}
\end{figure*}

The dots in \reffig{phasespace} show the discrete phase-space distribution of 
subhalos selected
according to the above criteria for each of the 9 template halos in our simulations. 
These distributions share broad similarity, but significant differences exist.
The similarity points to a basic dependence on the halo mass
while the differences reflect the halo-to-halo scatter that must be taken into
account in our method of halo mass determination. There appears to be a crude
relation between the dynamical status of subhalos and the formation history
of the host halo: the phase space is more extended and there are more unbound 
subhalos in late-formed halos such as A7 and A9 (Table \ref{tab:simu}).

\subsection{Observational Guidance} \label{sec:observation}

A practical application of the method presented in this paper is to
estimate the mass of the MW halo. We use the current observations
of the MW and its satellite galaxies as a guide in developing the
method. 

There are 13 satellite galaxies more luminous than $10^5\lsun$ 
within 300 kpc of the GC. Proper motions are available 
for 12 of these with the exception being Canes Venatici I
\citep{Pawlowski2013a}. We exclude Sextans due to the very 
large uncertainty in its proper motion. Canis Major and Sagittarius
are also excluded because they are so close to the GC that
they are experiencing strong tidal disruption. Consequently, 9
satellite galaxies of the MW can be used as subhalo tracers
at present. Among these, the Large Magellanic Cloud (LMC) 
at $50 \pm 2\ \kpc$ is the closest to the GC while Leo I at 
$258 \pm 15\ \kpc$ is the farthermost \citep{McConnachie2012}.

For developing our method to estimate the mass of a test halo
based on comparison of the kinematic properties of its subhalos
with the phase-space distributions of template halos, the most 
pertinent guidance provided by current observations of MW satellite
galaxies is the number of these tracers with sufficiently accurate
kinematic data and the typical uncertainties in such data.
We adopt the following fiducial values characteristic of current 
observations.
\begin{enumerate} 
	\item The number of tracers with adequate kinematic data is
		\begin{equation} N = 9. \end{equation}
	\item The relative uncertainty in the distance to the Sun in
	the Heliocentric Standard of Rest (HSR) is
		\begin{equation} (\sigma_r / r)_{\rm HSR} = 0.06. 
		\label{eq:err_r}\end{equation}
	\item The measurement error of the radial velocity with respect to 
	HSR is
		\begin{equation} 
		(\sigma_{\vlos})_{\rm HSR} = 1\ \kms. \label{eq:err_vlos}
		\end{equation}
	\item The precision for the proper motion components with 
	respect to HSR is
		\begin{equation} 
		(\sigma_{\mu_{\alpha}})_{\rm HSR} = 
		(\sigma_{\mu_{\delta}})_{\rm HSR} = 0.08\ \masyr.
		 \label{eq:err_pm}\end{equation}
\end{enumerate}

The uncertainties 
$(\sigma_r / r,\sigma_{\vlos}, \sigma_{\mu_{\alpha}},\sigma_{\mu_{\delta}})_{\rm HSR}$
adopted above are approximately the root-mean-square values of current 
precisions for luminous MW satellite galaxies \citep{Pawlowski2013a}. Note 
that the measurement errors of different observables are independent
as they are determined by separate methods and that the error 
in proper motion dominates. We will also check how our method 
is affected when different values from the fiducial ones are used
for the number of tracers and the measurement errors.

The measurement errors will be taken into account in 
constructing the phase-space distribution of subhalos for a
template halo. As will be described in detail in \refsec{mock}, 
this is done by making mock observations of kinematic properties 
of subhalos in a frame equivalent to HSR and then transforming 
the results into those with respect to the halo center acting as 
the GC. To make this transformation, we need the position and 
motion of the Sun in the Galactocentric Standard of Rest (GSR). 
We adopt the following distance and velocity of the Sun
relative to the GC \citep{Bland-Hawthorn2016a}:
\begin{equation}
\begin{split}
r_{\odot} &= 8.2 \pm 0.1\ \kpc, \\
(U_{\odot}, V_{\odot}', W_{\odot})_{\mathrm{GSR}} &= 
(10, 248, 7) \pm (1, 3, 0.5)\ \kms,
\end{split}
\label{eq:sun_gsr}
\end{equation}
where $U_{\odot}$ is the velocity towards the GC, 
$V_{\odot}'$ is positive in the direction of Galactic rotation, and
$W_{\odot}$ is positive towards the North Galactic Pole.
Note that $V_{\odot}'$ is the net rotation velocity of the Sun around the GC.

For simplicity, we drop the subscripts ``HSR'' and ``GSR'' below 
and note that 
$(\sigma_r / r,\sigma_{\vlos}, \sigma_{\mu_{\alpha}},\sigma_{\mu_{\delta}})$
always refer to HSR and $(U_{\odot}, V_{\odot}', W_{\odot})$ to GSR.


\subsection{Mock observations}\label{sec:mock}

While simulations yield precise values of $(r,v_r,v_t)$ for each subhalo,
we must take observational errors into account when comparing the 
corresponding phase-space distributions for template halos 
with the kinematic data on the observed satellite tracers to estimate 
the unknown mass of a test halo (see \refsec{method}). For developing
the method, we assume that all observed satellites have the same
measurement errors 
$(\sigma_r / r, \sigma_{\vlos}, \sigma_{\mu_{\alpha}},\sigma_{\mu_{\delta}})$.
We then make mock observations with these uncertainties in a frame
equivalent to HSR for all the selected subhalos of a template halo 
(see \refsec{subhalo_sample}). We also include the uncertainties in
the position and velocity of the Sun when transforming the mock data
in HSR into those with respect to the center of the template halo
that serves as the GC. The above procedure results in a smoothed
phase-space distribution of subhalos for the template halo and
accounts for the measurement errors at the same time.

We produce the mock data as follows.
\begin{itemize}
  \item Define ``GSR''

  We first apply a random rotation \citep{Arvo1992} to the simulations
  and require the center of a template halo to rest at the ``GC''. Perhaps a
  better practice is to adopt the orientation of the angular momentum of the
  inner halo as the ``Galactic North'' \citep[e.g.][]{Xue2008} instead of applying
  a random rotation. However, the difference would be very small as 
  the satellite tracers are far from the GC.
  
  \item Define ``HSR''

  We set the ``Sun'' at the point $(r_{\odot}, 0, 0)$ with velocities
  $(U_{\odot}, V_{\odot}', W_{\odot})$ in ``GSR''.
  To account for the uncertainties, we sample $r_{\odot}$ and 
  $(U_{\odot}, V_{\odot}', W_{\odot})$ from Gaussian distributions with
  means and standard deviations given in \refeq{sun_gsr}.
  
  \item Observe subhalos in ``HSR''
  
  We ``observe'' in ``HSR'' the distance, radial velocity, and proper motion
  of subhalos according to Gaussian distributions with standard deviations
  given in Equations (\ref{eq:err_r}), (\ref{eq:err_vlos}), and (\ref{eq:err_pm}),
  respectively. The measurement errors are taken to be independent of each
  other as they correspond to separate methods in real observations.
  
  \item Transform data from ``HSR'' to ``GSR''
    
  We convert the mock data in ``HSR'' into $(r,v_r,v_t)$ in ``GSR'' for each subhalo,
  which are then used to calculate the corresponding $(E,L)$. Note that in this step
  we adopt the central values of the ``solar'' position and velocities in ``GSR''.
  
  \end{itemize}

Following the above procedure, we make 2000 mock observations of
the $i$-th subhalo of a template halo and obtain the probability density 
$\tilde p (E, L| \mathrm{sub}_i)$ for this subhalo to be ``observed'' at $(E,L)$.
The quantity $\tilde p (E, L| \mathrm{sub}_i)$ can be approximated by a 2D 
Gaussian distribution
\begin{equation}
	\tilde p (E, L| \mathrm{sub}_i) = \frac{1}{2 \pi S \sqrt{1 - \rho^2}}
 \exp \left( - \frac{x^2 + y^2 - 2 \rho xy}{2 (1 - \rho^2)} \right),
\label{eq:obs_err}
\end{equation}
where
\begin{gather}
x = \frac{E - \mathrm{avg} (E_{i j})}{\sqrt{\mathrm{var} (E_{i j})}},\\
y = \frac{L - \mathrm{avg} (L_{i j})}{\sqrt{\mathrm{var} (L_{i j})}},\\
\rho = \frac{\mathrm{cov} (E_{i j}, L_{i j})}{\sqrt{\mathrm{var} (E_{i j}) \mathrm{var} (L_{i j})}},\\
S = \sqrt{\mathrm{var}(E_{i j}) \mathrm{var} (L_{i j})},
\end{gather}
and where $E_{i j}$ and $L_{i j}$ are the results from the $j$-th mock observation
of the $i$-th subhalo. In the above equations, the average, variance, and covariance
refer to operations on $j$ only. Note that due to the 
relatively large uncertainties in proper motion measurement, the variances 
cannot be estimated reliably by analytical error propagation.

\refeq{obs_err} shows that accounting for measurement errors through
mock observations turns a discrete point representing the $i$-th subhalo into 
a smooth distribution $\tilde p (E, L| \mathrm{sub}_i)$ in the phase space.
In this procedure, measurement errors also shift the mean position of the subhalo 
in the phase space from that given by the simulations.


\subsection{Constructed subhalo phase-space distribution} \label{sec:phase_space}

Repeating the procedure for obtaining the probability density 
$\tilde p (E, L| \mathrm{sub}_i)$ for all the selected subhalos in a template halo of mass $\mh$, 
we construct the corresponding phase-space distribution $p (E, L| \mh)$ as an average
of these probability densities [see \refeq{p(E,L)}]. In general, because the subhalo 
sample has a limited size, $p (E, L|\mh)$ may be discontinuous even after measurement
errors are taken into account. The discontinuity would be even more prominent 
were measurement errors to decrease significantly in the future.
Our method of halo mass determination requires a smooth $p (E, L|\mh)$,
which can be obtained conveniently by replacing $\mathrm{var} (E_{i j})$ and 
$\mathrm{var} (L_{i j})$ with $\widetilde{\mathrm{var}} (E_{i j}) = \mathrm{var} (E_{i j}) +S_E^2$
and $\widetilde{\mathrm{var}} (L_{i j}) = \mathrm{var} (L_{i j}) +S_L^2$, respectively, 
in \refeq{obs_err}. We take the smoothing terms to be
$S_E=\alpha E_h$ and $S_L=\alpha L_h$, where
$E_h=GM_h / R_h$ and $L_h = \sqrt{GM_h R_h}$ are the characteristic
energy and angular momentum per unit mass.
We add the smoothing adaptively, by choosing $\alpha$ for each subhalo
such that the ellipse with semiaxes of $S_E$ and $S_L$ covers just
the nearest 40 neighbors in the phase space.
We have checked that the result is not sensitive to the detailed choice of $\alpha$.
Using a fixed $\alpha = 0.1$--0.2 for all subhalos produces almost the same result.

\begin{figure*}[ht!]
\epsscale{1.2}
\plotone{\figdir/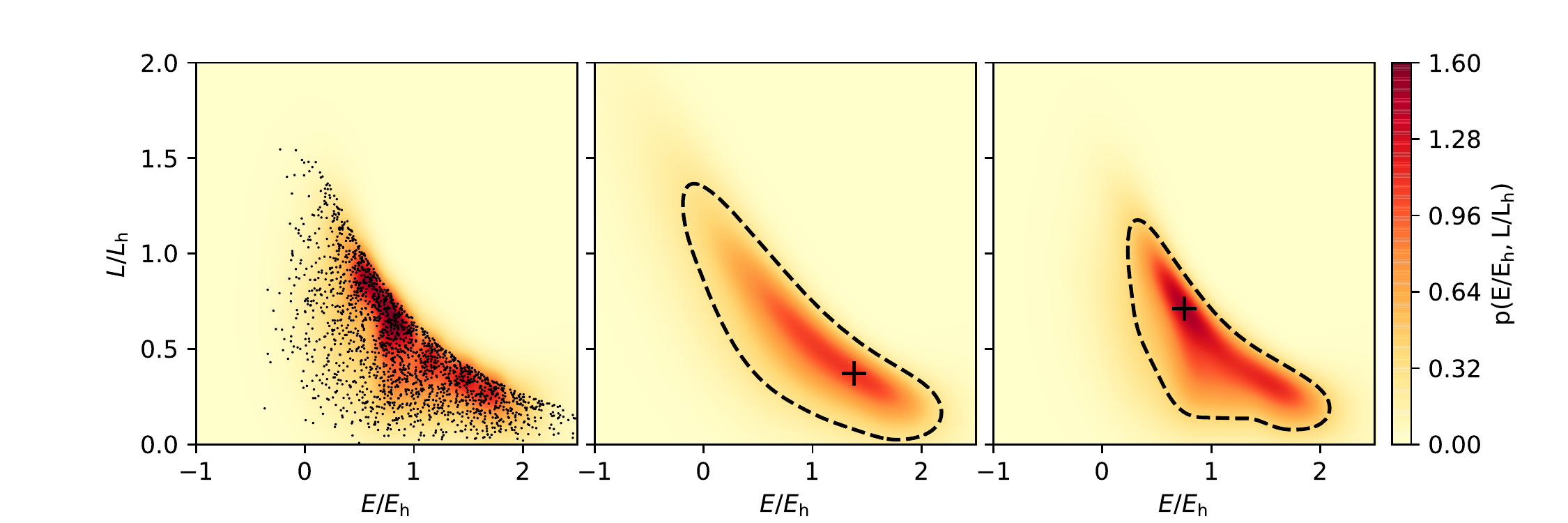}
\caption{Phase-space distribution of subhalos for template halo A1.
The discrete distribution in the left panel is taken directly from the simulations.
The smooth distribution in the middle panel includes the fiducial measurement
errors. The right panel assumes $\sigma_{\pmra}=\sigma_{\pmdec}=0.01\ \masyr$ instead of 
the fiducial values of $0.08\ \masyr$ while all other errors remain the same 
as for the middle panel. Color indicates the probability density and the cross 
marks the point of the highest density. 
The dashed curves enclose the densest 68\% of the region.
}
\label{fig:smooth}
\end{figure*}

\reffig{smooth} shows the phase-space distribution 
for template halo A1. Compared with the discrete distribution taken directly
from the simulations (left panel), the smooth distribution including the fiducial
measurement errors (middle panel) is more extended. The effects of the 
measurement errors are illustrated by the comparison of the middle and right panels.
For the latter, $\sigma_{\pmra}=\sigma_{\pmdec}=0.01\ \masyr$ are used instead of the fiducial values of 
$0.08\ \masyr$ while all other errors remain the same as for the middle panel.
The much smaller $\sigma_{\pmra}$ and $\sigma_{\pmdec}$ not only shrink the distribution, but also
shift the point of the highest probability density (marked as the cross).

\section{Tests with Mock Samples}\label{sec:test}

In our method of halo mass determination, we scale a template halo of
mass $\mh$ to different masses and obtain a family of subhalo phase-space
distributions $p(E,L|\mh')$ following the procedure presented in \refsec{template}. 
We then use these distributions and the $(E,L)$ data on the observed satellite 
tracers of a test halo to obtain the likelihood ${\cal L}({\rm obs}|\mh')$ as a function 
of $\mh'$ [see \refeq{likelihood}]. This gives the MLE $\mesti$ for the test halo mass
based on a specific set of scaled template halos.
If the test halo is the MW, we will use the actual kinematic data on 
its dwarf satellite galaxies. For each satellite, the actual measurement errors
will be used to make mock observations to obtain the corresponding 
$p(E,L|\mh')$ while the central values of the kinematic data in HSR will be
used (along with the central values of the solar position and velocities relative to
the GC) to obtain the corresponding $(E,L)$ in GSR. To test the validity and
accuracy of our method, we choose a subset from the subhalos of a template 
halo to serve as the ``observed'' satellite tracers. These tracers are referred to as 
the mock sample and their $(E,L)$ data are obtained by making a single mock 
observation of each tracer as described in \refsec{mock}. Below we present 
a series of tests of our method using these mock samples.

\subsection{Bias in the MLE for a specific template halo} \label{sec:test_one}

\begin{figure}[htb!]
\epsscale{1.2}
\plotone{\figdir/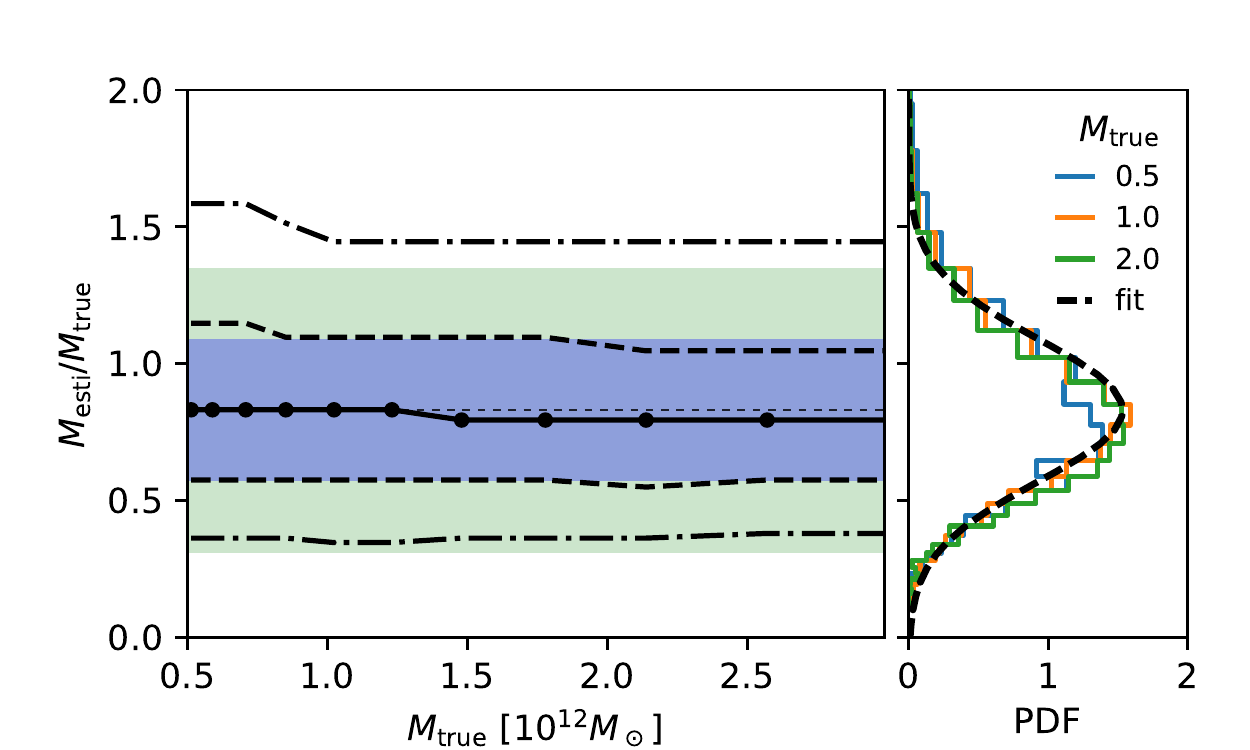}
\caption{
Distribution of $\mesti/\mtrue$ as a function of $\mtrue$ for test halos scaled from
template halo A1. Left panel: The solid, dashed, dot-dashed curves show the 
median value and the 68\% ($1\sigma$) and 95\% ($2\sigma$) intervals for 
$\mesti/\mtrue$ as functions of $\mtrue$, which agree very well with the
corresponding characteristics of the Gaussian distribution $\mathcal{N}(0.83, 0.26^2)$ 
(thin dotted line and shaded regions). 
Right panel: The histograms show the distributions of $\mesti/\mtrue$ for
$\mtrue=(0.5, 1, 2)\times 10^{12}\msun$, which are in excellent agreement with
the dashed curve showing the Gaussian distribution.
}
\label{fig:test1}
\end{figure}

We first check how well the true mass $\mtrue$ of a halo can be recovered by 
the MLE $\mesti$ in our method. As an example, we use template halo A1 as
both the test halo to generate mock samples and the template to estimate
the mass of the test halo.
We randomly pick 9 of its subhalos to make a mock sample and apply our
method to obtain the corresponding $\mesti$. We repeat this with 5000 random
mock samples to obtain a distribution of 
$\mesti/\mtrue$ at $\mtrue=1.58\times 10^{12}\msun$ for template halo A1.
We then use halos scaled from template halo A1 as test halos
and obtain distributions of
$\mesti/\mtrue$ for $\mtrue=(0.5$--$3.0)\times 10^{12}\msun$. We find that
to very good approximation, all these distributions can be fitted to a single 
Gaussian $\mathcal{N}(0.83, 0.26^2)$ with a mean of 0.83 and a standard 
deviation of 0.26. This is illustrated by the excellent agreement between
the histograms showing the distributions for 
$\mtrue=(0.5,1,2)\times 10^{12}\msun$ and the dashed curve for the 
Gaussian fit in the right panel of \reffig{test1}. In addition, the left panel
of this figure shows the median value (solid curve) and the 68\% ($1\sigma$, 
dashed curves) and 95\% ($2\sigma$, dot-dashed curves) intervals for 
$\mesti/\mtrue$ as functions of $\mtrue$. These again agree 
very well with the Gaussian fit.

As \reffig{test1} shows, the MLE $\mesti$ tends to underestimate the halo mass
$\mtrue$ with a bias that is nearly independent of $\mtrue$. 
Recall that the likelihood is constructed from the phase-space distribution 
$p(E,L|\mh')$ as a function of $\mh'$. Because $E$ also depends on $\mh'$,
the likelihood is non-Bayesian and gives a biased MLE. 
As shown in \reffig{scaling}, 
$p(E,L|\mh')$ is denser for a lower $\mh'$. Thus the likelihood tends to favor 
a smaller halo mass than the true value. This bias is intrinsic to our method, but
as shown below, it is insensitive to the number of tracers used, the measurement 
errors, or the formation history of the halo. Therefore, we can use
$\mcorr=\mesti/\bias$ with $\bias=\langle\mesti/\mtrue\rangle=0.83$ 
as an essentially unbiased estimator
for the halo mass. However, the relative uncertainty of $\mcorr$ depends on
the number of tracers used and the measurement errors, being $\sim 30\%$ 
for 9 tracers with the fiducial measurement errors.

\begin{figure*}[htb!]
\epsscale{1.2}
\plotone{\figdir/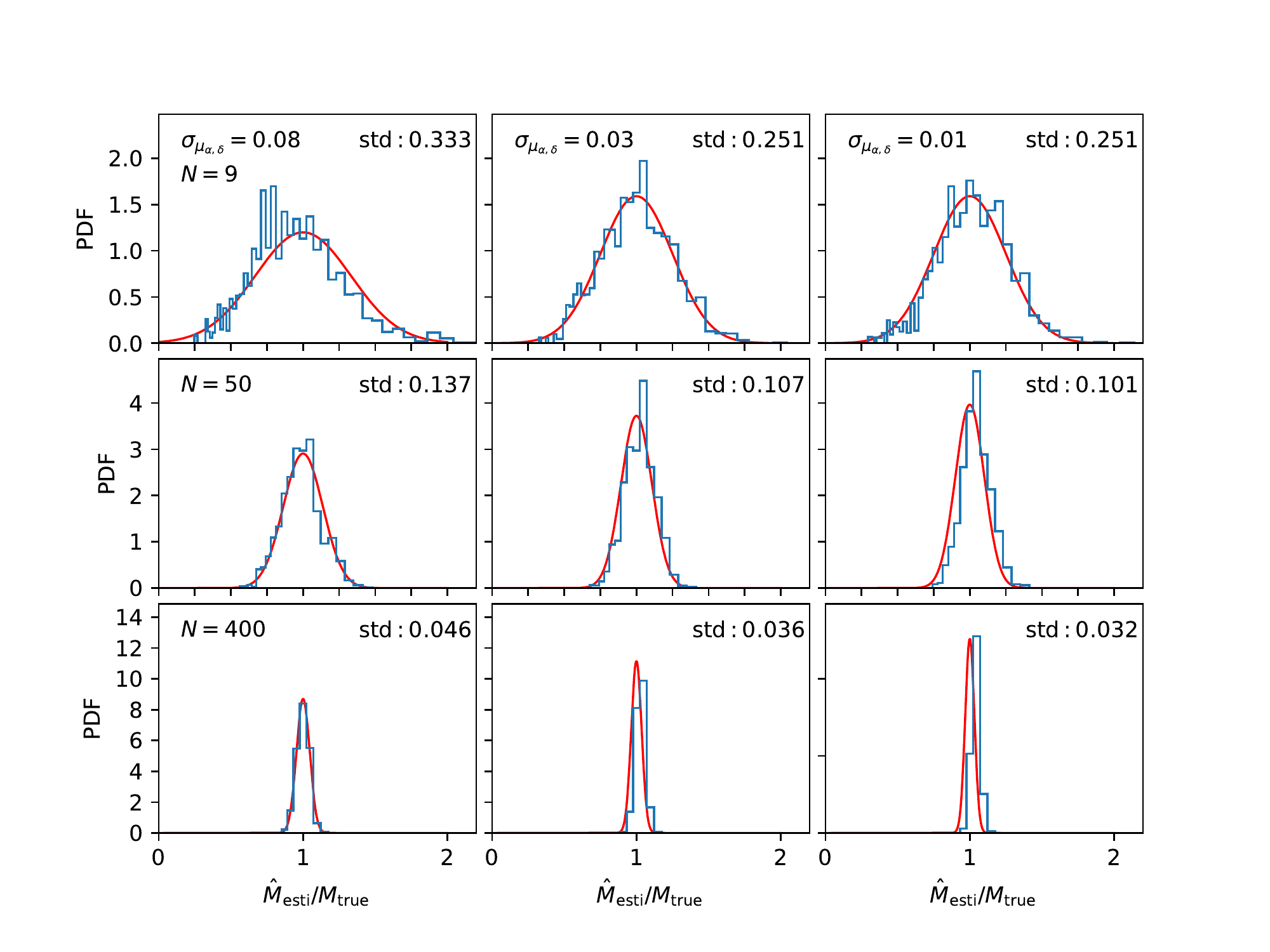}
\caption{
Distributions of $\mcorr/\mtrue$ for
different numbers ($N$) of tracers and observational errors
($\sigma_{\mu_{\alpha,\delta}}$) when template halo A1 is used
as the test halo. The upper, middle, and lower
rows assume $N=9$, 50, and 400, respectively. The left, middle,
and right columns assume $\sigma_{\mu_{\alpha,\delta}}=0.08$,
0.03, and 0.01 $\masyr$, respectively. In each case, the histogram
showing the distribution can be well described by the red curve 
showing a Gaussian centered at $\mcorr/\mtrue=1$ with the 
standard deviation (``std'') indicated in the corresponding panel.
}
\label{fig:test1_multi}
\end{figure*}

Using template halo A1 as the test halo, we show in
\reffig{test1_multi} the distributions of $\mcorr/\mtrue$
for different numbers ($N$) of tracers 
and observational errors. We take $N=9$, 50, and 400, respectively.
As proper-motion measurements dominate the observational uncertainties, 
we take $\sigma_{\mu_\alpha}=\sigma_{\mu_\delta}=0.08$, 0.03, and 0.01 $\masyr$, 
respectively, but keep the other measurement errors at their fiducial values.
For each choice of $(N,\sigma_{\mu_{\alpha,\delta}})$,
the distribution of $\mcorr/\mtrue$ (histogram) can be well fitted by a 
Gaussian (red curve) centered at $\mcorr/\mtrue=1$ with the standard deviation
(``std'') indicated in the corresponding panel of \reffig{test1_multi}.
This demonstrates that the fixed bias-correction $\eta=0.83$ works quite
well for very different numbers of tracers and measurement errors.
As shown in \reffig{test1_multi}, the standard deviation of $\mcorr/\mtrue$ 
decreases when more tracers and more precise observations are used. 
For fixed measurement errors, this quantity exhibits the expected
$1/\sqrt{N}$ dependence.

\begin{figure}[htb!]
\epsscale{1.2}
\plotone{\figdir/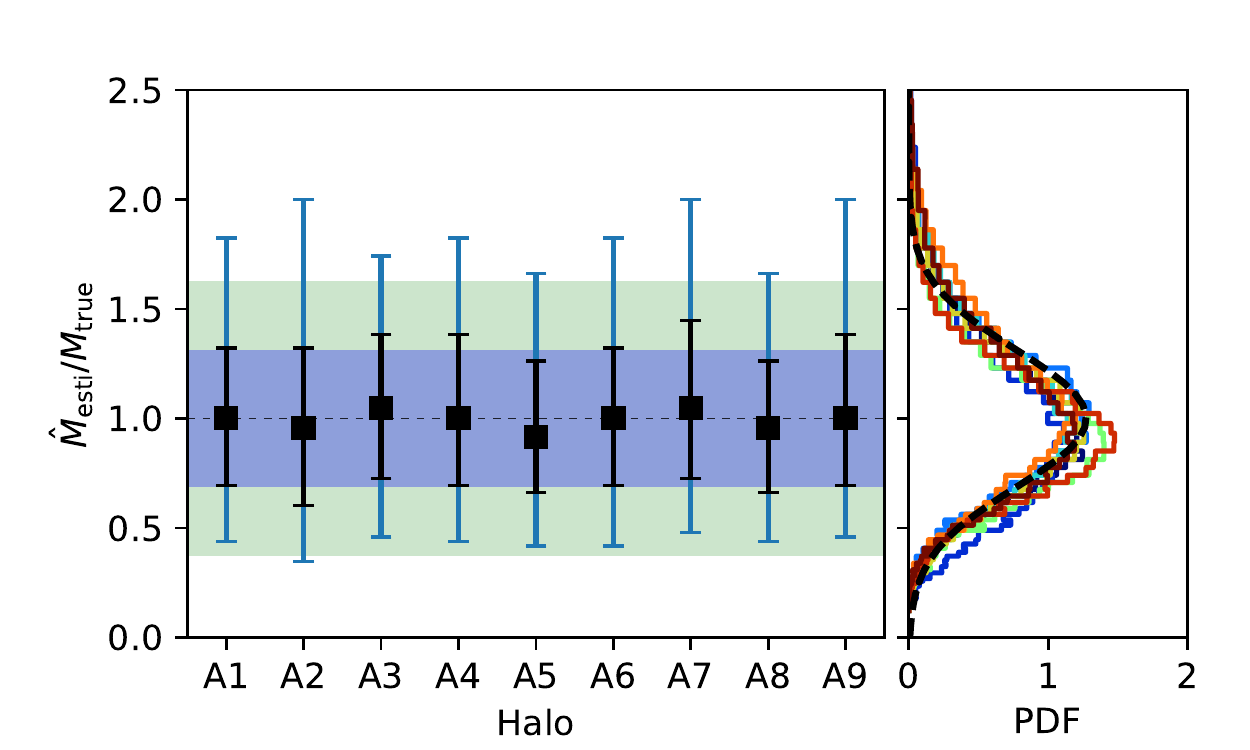}
\caption{
Distribution of $\mcorr/\mtrue$ when each of the 9 template halos is
used as the test halo.
Left panel: The filled squares and error bars show 
the median value and the 68\% ($1\sigma$) and 95\% ($2\sigma$) intervals 
for $\mcorr/\mtrue$. These compare very well with the thin dotted line and 
the shaded regions showing the corresponding characteristics of the Gaussian 
distribution $\mathcal{N}(1, 0.3^2)$. 
Right panel: The histograms showing the distributions of $\mcorr/\mtrue$ are
compared with the dashed curve showing the Gaussian distribution.
}
\label{fig:test2}
\end{figure}

We have also checked that the corrected halo-mass estimator $\mcorr=\mesti/\bias$ with
$\bias=0.83$ works consistently when each of the 9 template halos is used as both the 
test halo and the template to estimate the test halo mass. \reffig{test2} shows the distribution 
of $\mcorr / \mtrue$ in each case when the mock samples are observed with the fiducial 
measurement errors. All the distributions can again be well fitted by a Gaussian
$\mathcal{N}(1, 0.3^2)$. While the median value of $\mcorr / \mtrue$ fluctuates slightly 
around unity, this fluctuation is $< 5\%$, well within the standard deviation.
Therefore, $\mcorr$ recovers the true halo mass within $\sim 30\%$ for all the 9 template 
halos used as test halos. Because these halos have very different formation history 
and dynamical status, this demonstrates the validity and robustness of our method
at least when the formation history of a test halo is known.
 						
\subsection{Influence of subhalo mass} \label{sec:test_msub}

Our method implicitly assumes that the phase-space distribution of subhalos
is independent of their masses. This is supported by recent studies 
\citep[e.g.][]{Han2016}, which showed
that small and massive subhalos have very similar dynamics. This can be
understood because dynamical friction with strong mass dependence is 
important only for major mergers. Nevertheless, because the intended tracers
for the MW halo are its satellite galaxies, the more luminous of which tend to 
inhabit massive subhalos, we carry out further tests to check any possible 
influence of subhalo mass on our method.

\begin{figure}[htb!]
\epsscale{1.2}
\plotone{\figdir/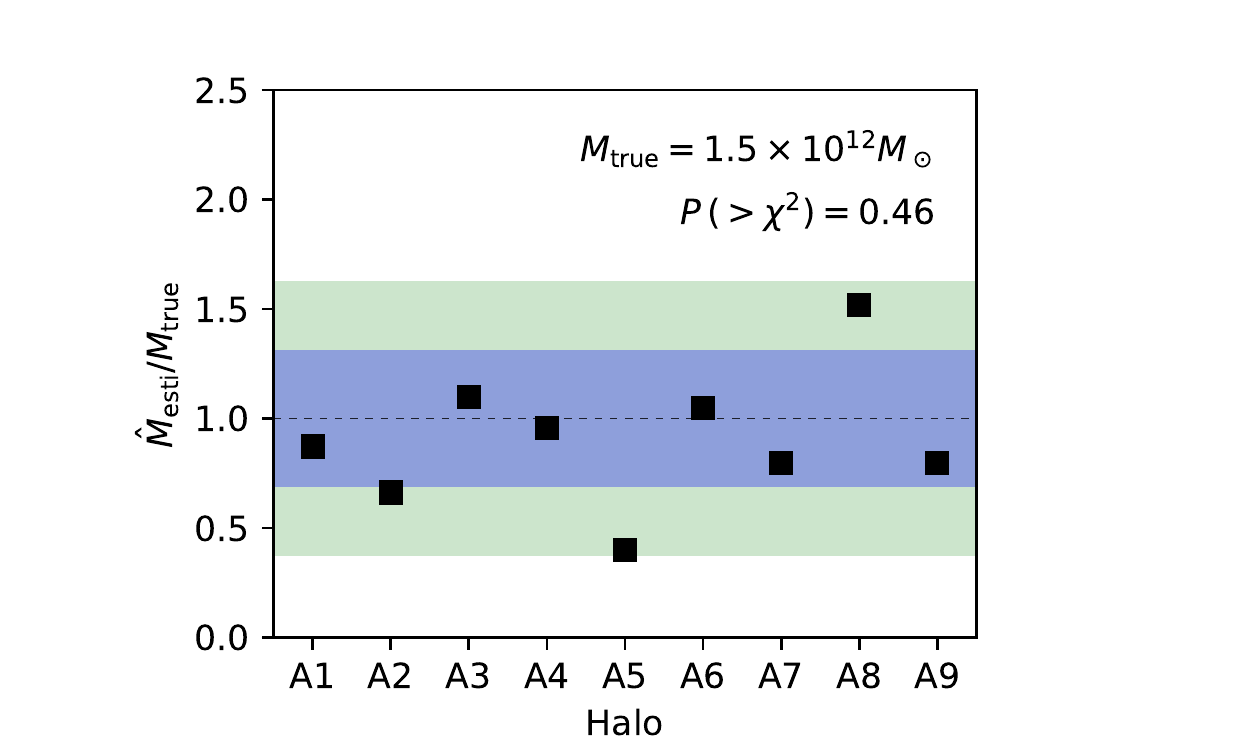}
\caption{
Results obtained from mock samples each having
the most massive 9 subhalos of a halo. The filled squares give the
$\mcorr/\mtrue$ when each of the template halos is scaled to 
$\mtrue=1.5\times10^{12}\msun$ and used as the test halo.
The thin dotted line and the shaded regions show the median and the
$1\sigma$ and $2\sigma$ intervals for the Gaussian distribution 
$\mathcal{N}(1,0.3^2)$. The p-value of the $\chi^2$ test of the filled
squares against the Gaussian distribution is $P(>\chi^2)=0.46$.
}
\label{fig:test2_max9}
\end{figure}

In the first test, we scale each of the 9 template halos to a mass of 
$\mtrue=1.5\times10^{12}\msun$ and take the most massive (at infall)  
9 subhalos in each case as the mock sample with the fiducial measurement 
errors. \reffig{test2_max9} shows $\mcorr/\mtrue$ (filled squares) for these 
test halos. These results are fully consistent with those in \reffig{test2}, which 
are obtained from mock samples each having 9 randomly-selected subhalos.
Specifically, when the results in \reffig{test2_max9} are compared with
the Gaussian distribution $\mathcal{N}(1, 0.3^2)$, the $\chi^2$ test gives 
$P(>\chi^2)=0.46$, which indicates no deviation. This insensitivity to the
mock sample is also confirmed when we scale the template 
halos to other masses and use those as test halos.

\begin{figure}[htb!]
\epsscale{1.2}
\plotone{\figdir/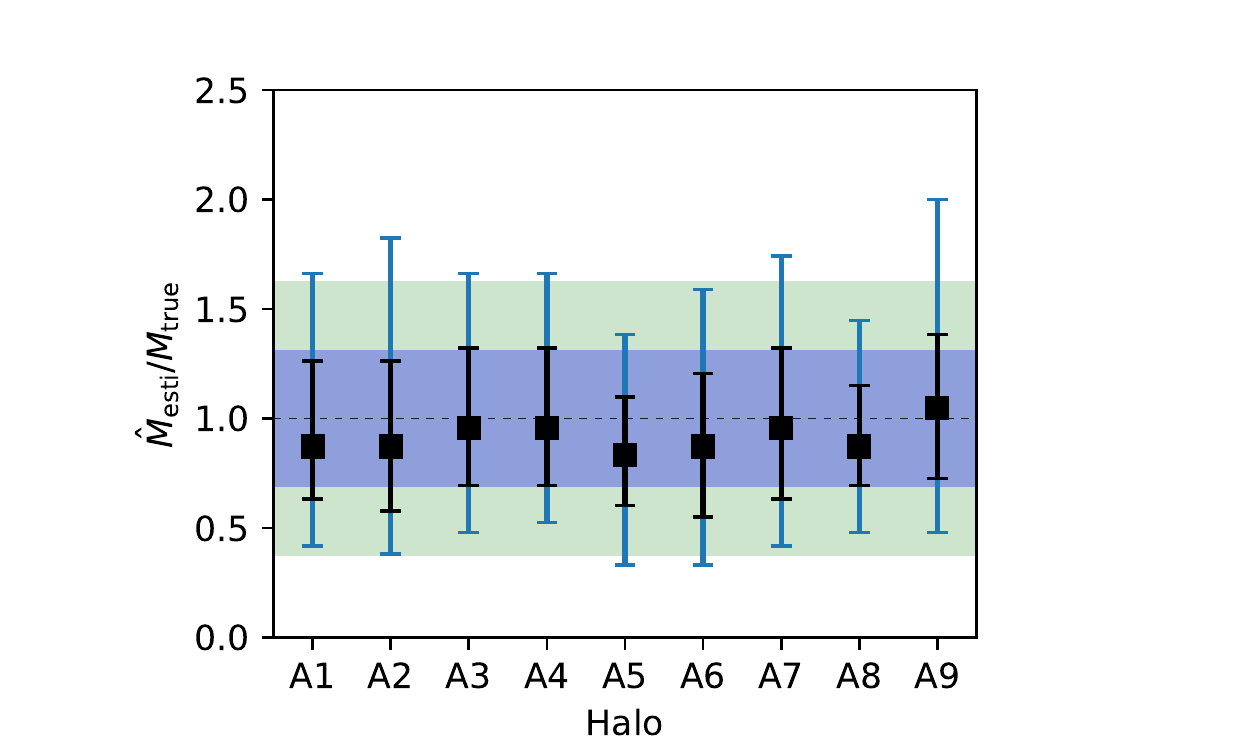}
\caption{
Same as the left panel of \reffig{test2}, but the mock samples are randomly 
drawn from the top 100 massive subhalos in each test halo. 
}
\label{fig:test2_max100}
\end{figure}

In the second test, mock samples are created by randomly selecting 9 subhalos
from the top 100 massive subhalos in each of the 9 template halos used as
test halos. We show 
in \reffig{test2_max100} the distribution of $\mcorr/\mtrue$ obtained from 5000 mock 
samples for each test halo. These results are almost the same as those in \reffig{test2}.

Based on the above tests and in view of the $\sim 30\%$ uncertainty
in our halo mass estimate, which is mostly due to the relatively small number of tracers 
used and the rather significant errors in proper motion measurement, we conclude that 
any influence of subhalo mass can be safely ignored. In any case, when the number of 
the MW satellite galaxies with precise kinematic data increases, this sample will reach 
tracers of lower luminosity associated with less massive subhalos, and therefore, 
become closer to a sample of randomly-selected subhalo tracers best suited for
our method.

\subsection{Halo-to-halo scatter} \label{sec:test_sys}

\begin{figure}[htb!]
\epsscale{1.2}
\plotone{\figdir/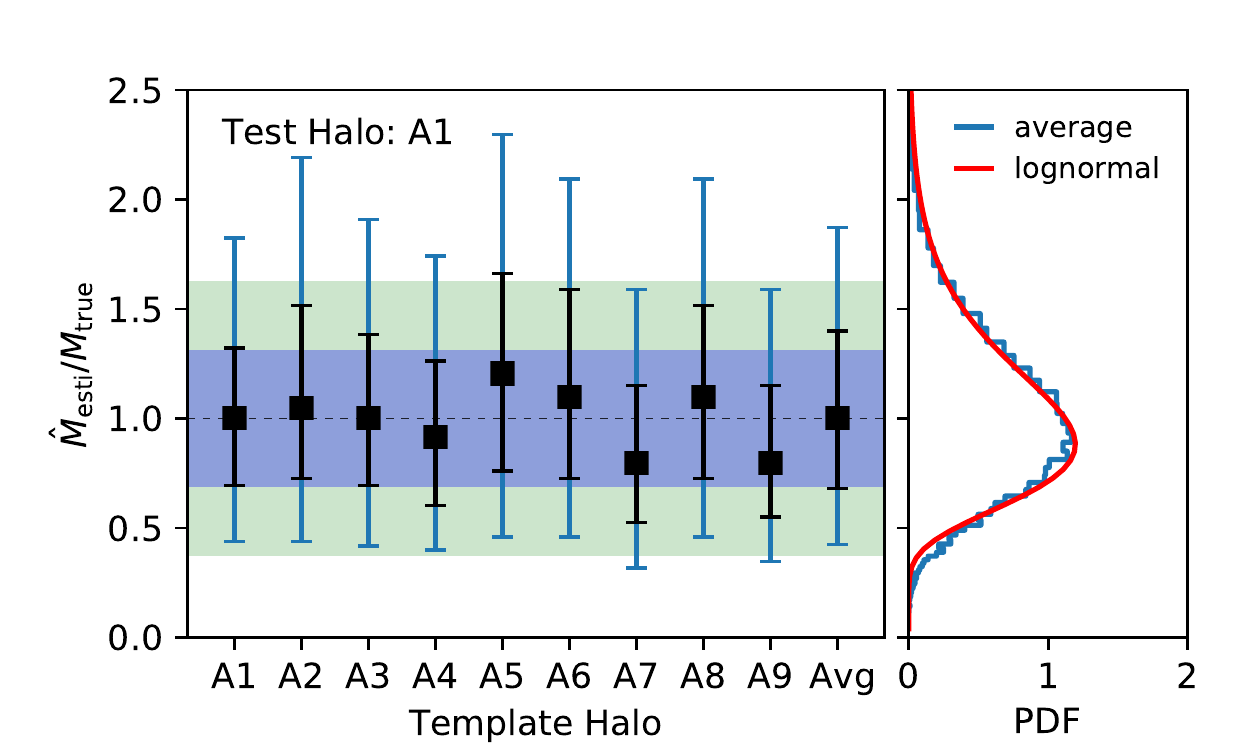}
\caption{
Distribution of $\mcorr/\mtrue$ obtained for test halo A1
by comparing mock samples from this halo with each of the 9 template halos.
Left panel: The filled squares and error bars show the median value
and the 68\% and 95\% intervals of $\mcorr/\mtrue$. The thin dotted line
and the shaded regions indicate a Gaussian distribution centered at unity
with a standard deviation of 0.3.
The rightmost filled square with error bars shows $\avgmcorr/\mtrue$ obtained 
by averaging the results from all 9 template halos. 
Right panel: The distribution of $\avgmcorr/\mtrue$ is compared with a lognormal 
$\ln \mathcal{N}(0, 0.35^2)$.
}
\label{fig:test3a1}
\end{figure}

So far we have shown that if the formation history of a test halo is known,
the halo mass can be determined reliably by our method.
However, halo formation history is not readily available in practice. 
Without such information, we must resort to comparing the kinematic
data on the observed tracers of a test halo with the subhalo phase-space
distributions for a number of template halos with a wide range of formation
history. For example, when we use template halo A1 as the test halo,
we estimate its mass using all the 9 template halos. For clarity, we
refer to template halo A1 in this case as test halo A1.
\reffig{test3a1} shows the distributions of $\mcorr/\mtrue$ for
test halo A1 obtained by comparing fiducial mock samples 
(9 tracers with fiducial measurement errors) from this halo
with the phase-space distribution from each of the 9 template halos.
The influence of halo formation history on $\mcorr/\mtrue$ can be seen
clearly. 

\begin{figure}[htb!]
\epsscale{1.2}
\plotone{\figdir/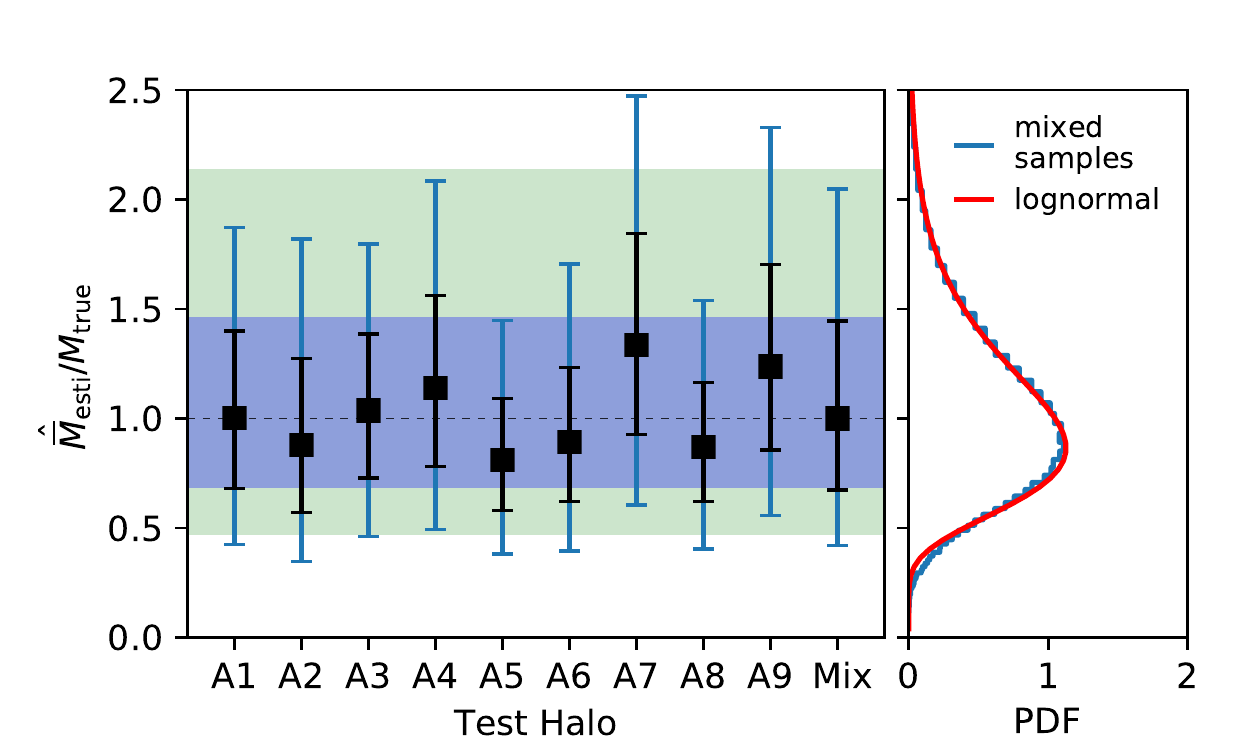}
\caption{
Distribution of $\avgmcorr/\mtrue$ for each test halo. 
Left panel: The filled squares and error bars show 
the median value and the 68\% and 95\% intervals for $\avgmcorr/\mtrue$.
The thin dotted line and the shaded regions show the median and the
$1\sigma$ and $2\sigma$ intervals for the lognormal distribution 
$\ln \mathcal{N}(0,0.38^2)$.
The rightmost filled square with error bars shows the result for the
mixed mock samples randomly drawn from the 9 test halos.
Right panel: The distribution of $\avgmcorr/\mtrue$ for the mixed mock 
samples is compared with the lognormal distribution.
}
\label{fig:test3}
\end{figure}

To address the lack of the formation history of a test halo, 
we take the average of the $\mcorr$ obtained for this halo
using the subhalo phase-space distribution for each of the template halos, 
\begin{equation}
\avgmcorr=\sum_{i=1}^{N_{\rm temp}}
\frac{\mcorr\phantom{}_{,i}}{N_{\rm temp}}, 
\label{eq:mavg}
\end{equation}
and check if this average (for $N_{\rm temp}=9$) gives a better estimate. 
We show the distribution of $\avgmcorr/\mtrue$ for test halo A1 in \reffig{test3a1}. 
The rightmost filled square in this figure shows that the median 
$\avgmcorr/\mtrue$ is 1.
On the other hand, the corresponding 68\% and 95\% intervals are
asymmetric and favor higher values. As shown in the right panel of
\reffig{test3a1}, the distribution of $\avgmcorr/\mtrue$ is well described
by a lognormal $\ln \mathcal{N}(0, 0.35^2)$.
Using other template halos as test halos, we show in \reffig{test3}
the corresponding distributions of $\avgmcorr/\mtrue$. The filled
squares with error bars in the left panel show the median value and
the 68\% and 95\% intervals of $\avgmcorr/\mtrue$ for each test halo.
The median $\avgmcorr/\mtrue$ scatters around unity within 
$\sim 20\%$ ($\sim 30\%$ for a few cases), reflecting the difference
in formation history between the test and template halos.
We have also checked that the bias of $\avgmcorr/\mtrue$
does not depend on the number of tracers used or the measurement 
errors (not shown). Note that the relative uncertainty in 
$\avgmcorr/\mtrue$ is $\sim 30\%$ for all 9 test halos.

As a final test, we make mock samples by randomly picking a test halo
and then randomly selecting 9 subhalos from this halo. Based on 5000
such mixed mock samples, we show the distribution of $\avgmcorr/\mtrue$
as the rightmost filled square with error bars (left panel) and the histogram
(right panel) in \reffig{test3}. This distribution best characterizes the halo mass
estimate given by our method in practice, and is well described by a lognormal 
$\ln \mathcal{N}(0, 0.38^2)$, whose $1\sigma$ interval corresponds to
the interval $[0.68, 1.46]$ for $\avgmcorr/\mtrue$. The uncertainty is slightly 
larger than the case with mock samples from a single test halo
because a discrepancy of $\sim 20\%$ due to halo-to-halo scatter is also
included in addition to the statistical uncertainty. More precisely, the uncertainty 
due to halo-to-halo scatter is $19_{-4}^{+6}\%$ based on the variance of the 9 
data points in \reffig{test3}. 
While this uncertainty is irreducible without additoinal information,
it can be estimated better with more test halos.
In addition, the limited number (9) of template halos also introduces an
uncertainty of $\sim 20\%/\sqrt{9}\approx 7\%$ in $\avgmcorr/\mtrue$.
However, this uncertainty is relatively small compared to that from halo-to-halo scatter.

In principle, knowledge on the formation history
of a test halo can reduce the uncertainty in its mass estimate due to
halo-to-halo scatter. Of particular importance is information on the growth 
of the halo potential as well as the accretion and disruption of substructures.
However, it is difficult to find a simple indicator to 
characterize the influence of the halo assembly history
on the kinematics of surviving substructures. We intend to study
this problem in the future.


\subsection{Prospects and limitation} \label{sec:test_err}

Based on the preceding discussion, there are two main sources of uncertainties
in our method of halo mass determination: one is statistical and due to the 
limited number of tracers and measurement errors, while the other is
intrinsic and due to the lack of knowledge about the formation history of a test
halo. Below we quantify these uncertainties using mixed mock samples created
by randomly picking one of the test halos and then randomly selecting
a subset of its subhalos. We vary the sample size (the number $N$ of
tracers) and the error $\sigma_{\mu}= \sigma_{\mu_{\alpha}} = \sigma_{\mu_{\delta}}$
in proper motion measurement, which dominates the observational uncertainties.
The other measurement errors are kept at their fiducial values.

Because $\avgmcorr/\mtrue$ follows a lognormal distribution,
we define the uncertainty of our method as
\begin{equation}
 \sigma = \sqrt{\mathrm{var}\left[\ln\left(\avgmcorr/\mtrue\right)\right]}\,.
\label{eq:sigma}
\end{equation}
Note $\mathrm{var}(X) \simeq \mathrm{var}(\ln X)$ when $X$ follows $\ln \mathcal{N}(0, \sigma^2)$ and $\sigma$ is small.

\begin{figure}[htb!]
\epsscale{1.2}
\plotone{\figdir/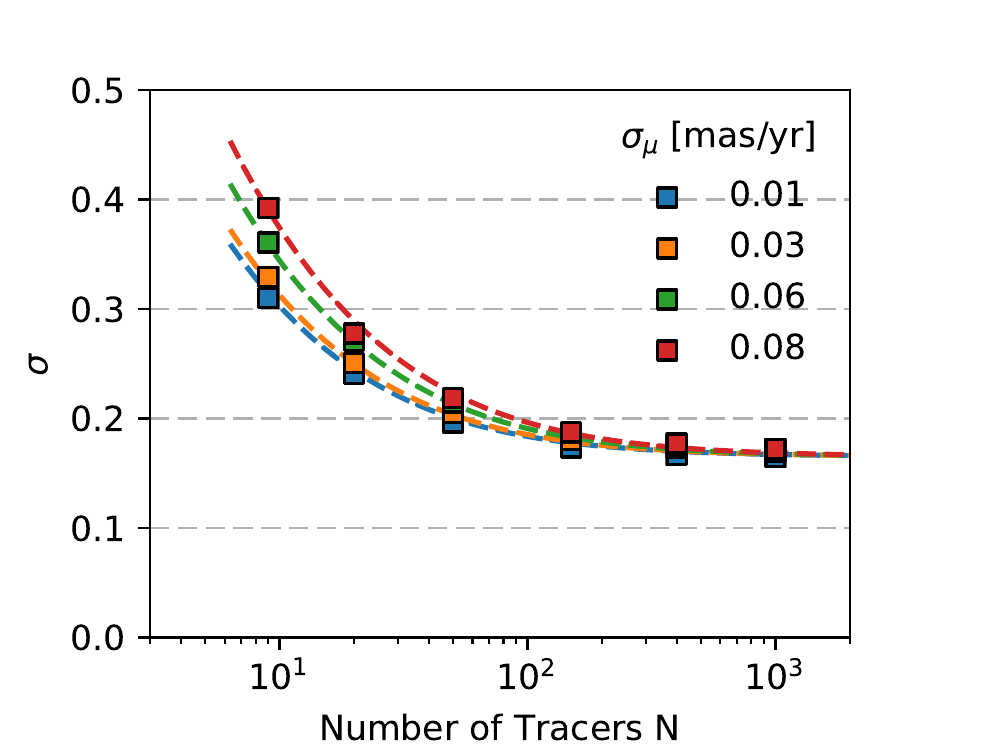}
\caption{Uncertainty $\sigma$ in $\ln\left(\avgmcorr/\mtrue\right)$ 
as a function of the number $N$ of tracers for different values of the error 
$\sigma_\mu$ in proper motion measurement. 
Filled squares are data obtained using mixed mock samples randomly 
drawn from the 9 test halos. Good fit to the data is provided by the dashed 
curves for $\sigma^2=(8.75^2\sigma_\mu^2+0.80^2)/N+0.17^2$.}
\label{fig:error}
\end{figure}

In Figure \ref{fig:error}, we show $\sigma$ as a function of the number $N$ of
tracers for $\sigma_{\mu}=0.01$, 0.03, 0.06, and 0.08\ $\masyr$, respectively.
As shown by the dashed curves, these results can be well described by
\begin{equation}
\sigma^2 = \sigma_{\mathrm{stat}}^2 + \sigma_{\mathrm{hist}}^2 = \frac{A^2 \sigma_{\mu}^2 + \sigma_{\mathrm{other}}^2}{N} + \sigma_{\mathrm{hist}}^2,
\end{equation}
where $\sigma_{\mathrm{stat}}$ is the statistical term that decreases
with increasing $N$, and $\sigma_{\mathrm{hist}}$ is the intrinsic term 
due to the lack of knowledge about the halo formation history.
The statistical term can be specified by $A\sigma_{\mu}$, where
$A$ is a constant, and $\sigma_{\mathrm{other}}$, which captures the
other observational uncertainties and errors in constructing the subhalo
phase-space distribution. Fit to the data gives $A=8.75$, 
$\sigma_{\mathrm{other}}=0.80$, and $\sigma_{\mathrm{hist}}=0.17$.
Note that $\sigma_{\mathrm{hist}}$ depends on the template halos
used and may be estimated better with more halos in addition to 
the 9 used here. 

For the fiducial number of tracers ($N=9$) with the fiducial measurement errors
($\sigma_\mu=0.08\ \masyr$), $\sigma$ is $\sim 40\%$. If $N$ increases to 30,
$\sigma$ decreases to $\sim$25\% for the fiducial measurement precision.
However, as $N$ increases further, 
$\sigma$ becomes dominated by $\sigma_{\mathrm{hist}}$.
This sets a limiting number of tracers at $N\sim 50$, beyond which there is no 
significant gain in the accuracy of our halo mass estimate. This is similar to
the result of \citet{Wang2016b}, who gave a systematic uncertainty of 25--40\% 
for the MW mass estimate using dynamical tracers under the steady-state 
assumption. We emphasize that the ultimate improvement of our method
requires detailed knowledge about the formation history of a test halo.


\section{Conclusions}\label{sec:conclusion}

We have presented a method to estimate the mass of a dark matter halo
using the kinematic data of its subhalo tracers, which are satellite galaxies 
in practice. The halo mass is inferred by comparing these data with the 
distribution in the phase space of binding energy and angular momentum 
for subhalos in each of the template halos obtained in cosmological 
simulations. We have tested the validity and accuracy of this 
method with mock samples and found that the halo mass can be recovered 
within $\sim 40\%$ by using 9 tracers with the current observational precision. 
The uncertainty can be reduced to $\sim$25\% if the number of tracers with 
sufficiently accurate proper motion measurement increases to 30 in the future.
However, the subhalo phase-space distribution depends on the halo formation 
history and the lack of this knowledge results in an intrinsic uncertainty of 
$\sim 20$\% in our halo mass estimate, which cannot be reduced by increasing 
the number of tracers. 
Further studies on the assembly history of a halo 
and how this history affects the kinematics of its substructures are
essential to an accurate determination of its mass.
A direct application of our method is to estimate the mass of the MW halo.
Using the data on its 9 dwarf satellite galaxies, we obtain a mass of
$1.3\times 10^{12}\msun$ with uncertainties comparable to the expected value of
$\sim 40\%$. This preliminary result is consistent with various estimates in the
literature. A detailed report will be given elsewhere.

Although they do not seem to affect our current results, several issues 
regarding our approach merit discussion. We have found that the phase-space
distribution is nearly independent of subhalo mass. Because satellite galaxies
are the intended subhalo tracers, it is desirable to confirm this with further tests 
using satellite samples from semi-analytical or hydrodynamic simulations.
We have simulated 9 template halos with a wide range of formation history.
It is valuable to have more template halos to check if this range is sufficiently 
representative. Because high-resolution zoom-in simulations are required to 
provide well-resolved substructures for constructing the phase-space distributions,
it is computationally intensive to study many template halos.
Another issue is the influence of massive neighbors such as M31 in the case of 
the MW. Our 9 template halos are chosen to be relatively isolated to exclude 
such neighbors. Using a larger halo sample, we have checked that
the presence of a massive neighbor will not affect our method when
the distance to the neighbor exceeds three times its virial radius as in the case
of M31 and the MW (see Appendix).
Finally, in our mock observations, we set the origin of the ``GSR'' to rest at 
the center of a template halo. However, theoretical 
and observational studies suggest that central galaxies do not necessarily rest
at the centers of their host halos \citep[e.g.][]{Berlind2003,Yoshikawa2003}. 
A recent study by \cite{Guo2015a} reported that a central galaxy tends to 
move around the host halo center with a dispersion of 
$0.2 \sigma_{v, \mathrm{DM}}$ ($\sim 30\ \kms$ for the MW) for each velocity 
component. In addition, if the LMC exceeds 10\% of the MW mass, then 
the MW is moving relative to their barycenter at a velocity of $\sim 30\ \kms$ 
($v_{\mathrm{LMC}} \simeq 300\ \kms$ relative to the GC). In principle,
the unknown velocity offset between the GC and the MW halo center introduces 
an extra uncertainty in the MW halo mass estimate. However, in practice,
we find with Monte Carlo experiments that adding an extra velocity of 30 $\kms$ 
to the ``GC'' in mock observations only changes the results at the 
$\lesssim3\%$ level. So this effect might become significant only when the intrinsic
uncertainty in our method is reduced with information on the halo formation history.

Currently, our method is still limited by the number of tracers and 
measurement errors. Proper motions are only available for 12 of the 13 MW
satellite galaxies (the exception being Canes Venatici I) that are more luminous 
than $10^5\lsun$ and within 300 kpc of the GC. The best of these proper motion
measurements were made with HST. The Gaia mission will reduce uncertainties in 
proper motions of nearby classical satellites \citep[e.g.][]{vanderMarel2016} and 
make new measurements for fainter objects within $\sim$100 kpc of the Sun 
\citep{Wilkinson1999}. Proper motions of more distant satellite galaxies could be 
measured by a multi-year HST program with followup by the James Webb Space 
Telescope (JWST) or the Wide-Field Infrared Survey Telescope (WFIRST) 
\citep{Kallivayalil2015}. In addition, ongoing deep, wide-field sky surveys, 
such as the Dark Energy Survey (DES), PanSTARRS 1 (PS1), and VST ATLAS 
have doubled the number of known MW satellites over the past two years
\citep[e.g][]{Bechtol2015,Koposov2015,Drlica-Wagner2015,Laevens2015b,Torrealba2016a}.
The number of satellites brighter than the faintest known dwarf galaxies might eventually 
reach 300--600 and possibly as high as $\sim$1000 \citep{Tollerud2008}.
The above exciting progress in observations will undoubtedly enable us to 
determine the MW halo mass with increasing accuracy.


\acknowledgments
We thank Hui-Yuan Wang and You-Cai Zhang for their help in carrying out the N-body simulations, 
and Jia-Xin Han and Yang Wang for their help with identification of subhalos. 
We also thank the anonymous referee for constructive criticisms and helpful suggestions.
ZZL is grateful to Zheng-Yi Shao, Lu Li, Ying Zu, and Jia-Xin Han for helpful discussions of 
statistical methods. This work was supported in part by the NSFC
(11222325, 11320101002, 11533006, \& 11621303), 
the Knowledge Innovation Program of CAS (KJCX2-EW-J01),
973 Program No. 2015CB857003, 
Shanghai Key Laboratory Grant No.11DZ2260700, 
Shanghai talent development funding No. 2011069, and the US DOE (DE-FG02-87ER40328).

This work made use of the High Performance Computing Resource 
in the Core Facility for Advanced Research Computing at Shanghai Astronomical Observatory.


\software{Astropy \citep{AstropyCollaboration2013}, Gadget-2 \citep{Springel2005}}


\appendix \label{appendix}
\section*{Validation with more test halos from cosmological simulations}

\begin{figure}[htb!]
\epsscale{0.9}
\plotone{\figdir/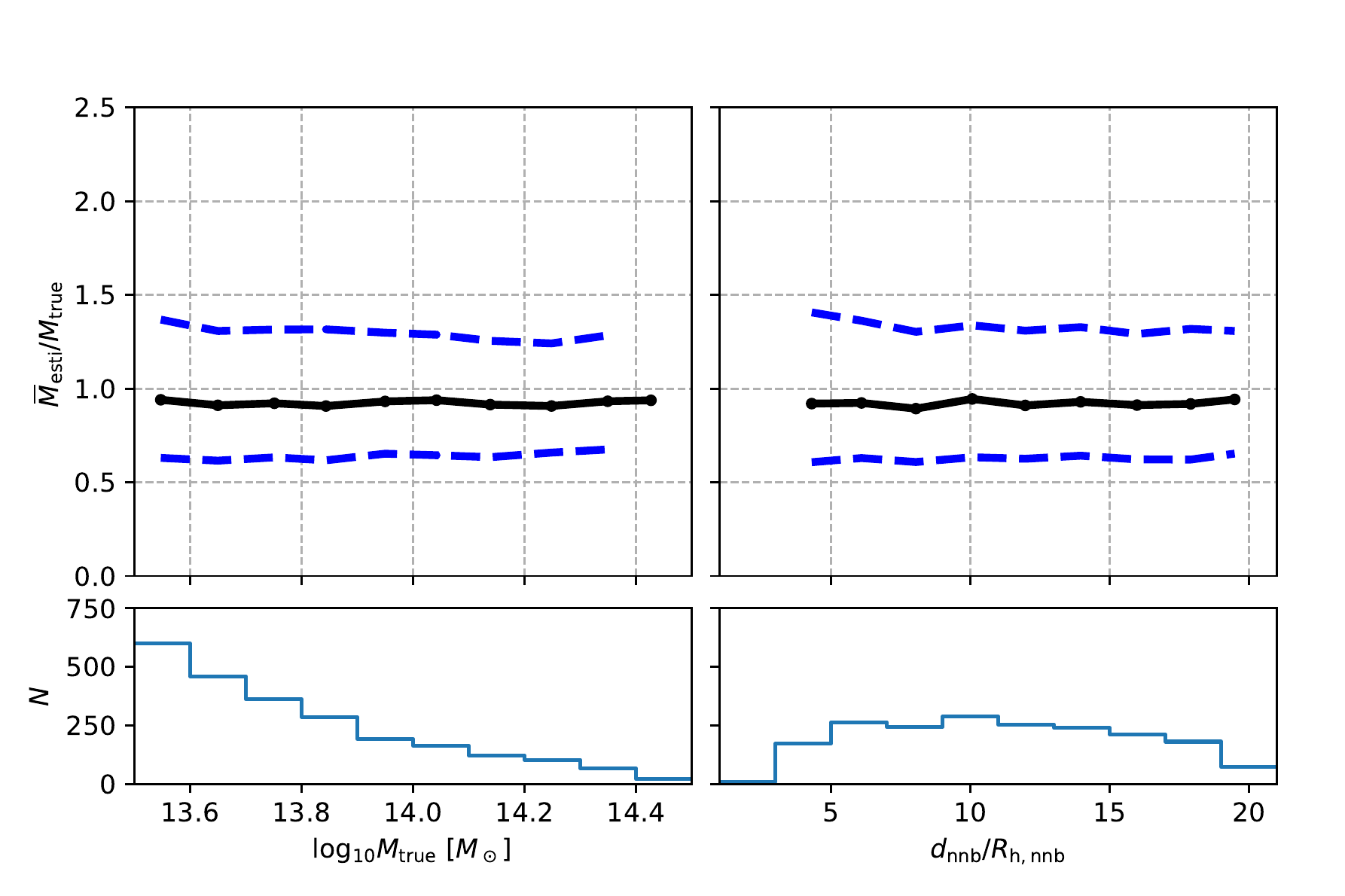}
\caption{
Distribution of $\avgmesti/\mtrue$ (upper panels) and
number of test halos $N$ (lower panels) as functions of the true
halo mass $\mtrue$ (left panels) and the distance to the nearest more 
massive neighbor relative to its halo radius
$d_{\rm nnb}/R_{\rm h, nnb}$ (right panels)
for a sample of test halos from cosmological simulations.
The solid and dashed curves in upper panels show the
median value and the 68\% ($1\sigma$) intervals for 
$\avgmesti/\mtrue$ in a bin of $\mtrue$ (left panel) and
$d_{\rm nnb}/R_{\rm h, nnb}$ (right panel). The histograms
in lower panels show the number of test halos in each bin.
}
\label{fig:test_cosmo_sample}
\end{figure}

Our method recovers the true halo mass consistently
in a series of tests with the 9 halos from zoom-in simulations. Nevertheless,
it is important to check the robustness of the method with a larger test halo sample.
Such a sample can also be used to investigate how a massive neighbor may 
affect the mass estimate for a halo, thereby checking the validity of using relatively
isolated template halos in our method. We select a set of test halos from 6 cosmological simulations.
Each simulation was performed with $1024^3$ particles in a periodic cubic box of 
$(150 \mpch)^3$. A $\LCDM$ cosmology was adopted with $\Omega_m = 0.268$, 
$\Omega_{\Lambda} = 0.732$, $h= 0.71$, $n_s=1$, and $\sigma_8 = 0.85$ \citep{Jing2007}.
The particle mass in these simulations is $2.3 \times 10^8 \msunh$.
We identify halos and subhalos in the same way as described in \refsec{simu}.
We note that the cosmological parameters adopted above are somewhat different 
from those in the main text. We expect that the effect of this difference would be small
compared to that of the difference in halo formation history for application of our method.

To ensure a sufficient number of well-resolved subhalos in each test halo,
we focus on halos in the mass range of $\sim2\times (10^{13}$--$10^{14})\msunh$
and obtain a sample of 2681 test halos. For each test halo, we select subhalos with
a maximum binding mass of $m_{\rm max}>300$ particles over history, a mass of
$m_0 > 10$ particles at present, and a distance of $160\ \kpc<r<1200\ \kpc$ to the
halo center. We then randomly pick 9 subhalos to make a mock sample.
(For the least massive halos, which amount to $< 1.5\%$ of the test halo sample and
contain fewer than 9 usable subhalos each, we randomly pick 9 subhalos with repetition.)
Mock observations of the 9 subhalos are made with measurement errors
$\sigma_r / r = 0.06$, $\sigma_{\vlos} = 4\ \kms$, and $\sigma_{\mu}=0.08\ \masyr$.
The above numerical values are the same as in \refsec{template}, except that
the distance range and $\sigma_{\vlos}$ are increased in proportion to the mass range 
of the test halos under consideration. Following the procedure in \refsec{template}, 
we estimate the mass of a test halo by comparing the phase space distribution of its 
mock sample with those of a template halo scaled to the same mass range.
The results using the 9 template halos A1--A9 are averaged to calculate the final
estimate $\avgmesti$. 

A total of 20 mock samples are chosen from each test halo to generate a distribution of 
$\avgmesti/\mtrue$, where $\mtrue$ is the true halo mass. This distribution is shown
in the upper left panel of \reffig{test_cosmo_sample} as a function of $\mtrue$ for the 
entire test halo sample. The solid and dashed curves give the median value and the 
68\% ($1\sigma$) intervals, respectively, for the $\avgmesti/\mtrue$ of all the mock 
samples in a bin of $\mtrue$. The number of test halos in each bin is shown in the 
lower left panel of \reffig{test_cosmo_sample}. It can be seen that although
the test halos [$\sim3\times(10^{13}$--$10^{14}) \msun$] have very different masses
from the template halos ($\sim1.5\times10^{12} \msun$), our method still gives reasonable 
mass estimates. The median value of $\avgmesti/\mtrue$ is nearly independent of
$\mtrue$ and can be taken as the bias $\eta=0.92$. In addition, the uncertainty of
$\avgmesti/\mtrue$ also has little dependence on $\mtrue$. These results are
similar to those shown in \reffig{test1} for MW-like test halos and demonstrate that 
our method using scaled templates is valid for a fixed range with a factor of $\sim 6$
variation in the halo mass even when these masses differ greatly from those for the 
template halos.

The distribution of $\avgmesti/\mtrue$ shown in \reffig{test_cosmo_sample} can be
well described by a lognormal with $\sigma=0.4$ [see similar definition in 
\refeq{sigma}], which extends the results in \refsec{test_err} to much larger
halo masses. However, the bias $\eta=0.92$ is $\approx 10\%$ larger than
the value of 0.83 adopted for MW-like halos. This small difference may be due to 
the difference in assembly history between MW-like halos and the test halos
under consideration, with possibly a minor contribution from the somewhat 
different cosmology adopted for the template and test halos. 
It is also possible that the template halos A1--A9 are not representative 
enough. While we cannot identify the exact cause for the above difference in $\eta$,
we note that this issue is secondary compared to the $\sim 40\%$ overall uncertainty 
of our method when applied to the MW with the current observational constraints. 
However, in anticipation of major improvement of observations, more detailed 
investigation with many more template halos is required to better
understand the influence of halo assembly history on our method.

For each of our template halos, its distance to any more massive halo 
exceeds three times the sum of the virial radii of both halos. We now
investigate how this choice of relatively isolated template halos may
influence the mass estimate. Satellites of a halo are subject to both
the gravitational force of the halo and the tidal force of a massive 
neighbor. For a halo of mass $\mh$ and virial radius $\rh$ with 
a neighbor of mass $M_{\rm nb}$ and virial radius $R_{\rm h,nb}$
at a distance $d_{\rm nb}$, the gravitational force of the halo on
a satellite of mass $m_{\rm sat}$ is $F_\textrm{g} \sim G\mh m_{\rm sat}/\rh^2$,
while the tidal force of the neighbor is 
$F_{\rm t}\sim GM_{\rm nb}m_{\rm sat}\rh/d_{\rm nb}^3$. So the importance
of the neighbor can be gauged by
$F_{\rm t}/F_{\rm g}\sim (M_{\rm nb}/\mh)(\rh/d_{\rm nb})^3$.
Because the halo and its neighbor have the same average density,
$M_{\rm nb}/R_{\rm h,nb}^3=\mh/\rh^3$, we obtain
$F_{\rm t}/F_{\rm g}\sim(R_{\rm h,nb}/d_{\rm nb})^3$.
Using the sample of test halos from cosmological simulations,
we locate every more massive halo in the neighborhood of a test halo
and define the one with the smallest $d_{\rm nb}/R_{\rm h,nb}$ as the nearest neighbor.
As shown in the upper right panel of \reffig{test_cosmo_sample},
the mass estimate is essentially independent of $d_{\rm nnb}/R_{\rm h, nnb}$
for $d_{\rm nnb}/R_{\rm h, nnb}\gtrsim 3$. Therefore, it is appropriate to use our
template halos with $d_{\rm nnb}/(R_{\rm h, nnb}+\rh)>3$ to estimate the masses of
those halos with $d_{\rm nnb}/R_{\rm h, nnb}\gtrsim 3$.
The galaxy M31 is perhaps slightly more massive than the MW and
is at a distance of $d_{\rm M31} \sim 800\ \kpc$. The virial radius of its 
dark matter halo can be estimated as $R_{\rm M31}\sim 200\ \kpc$. 
With $d_{\rm M31}/R_{\rm M31}\sim 4$, the effect of the tidal force of M31
is at the level of $\sim 2\%$ and our method can be safely applied
to estimate the mass of the MW halo.


\bibliography{MW_Mass}



\end{document}